\documentclass[aps,twocolumn,superscriptaddress,nofootinbib,longbibliography,notitlepage,floatfix]{revtex4-2}

\usepackage{amsmath,mathtools,amstext,amssymb,amsbsy,amsthm,mathrsfs,dsfont}
\usepackage{physics}
\usepackage{bm}
\usepackage{esint}
\usepackage{xfrac}

\usepackage{graphicx}
\usepackage[usenames,dvipsnames]{xcolor}
\usepackage{outlines}
\usepackage{subfigure}
\usepackage{diagbox}
\usepackage{tabularx}
\usepackage{longtable,booktabs,multirow,dcolumn,makecell}
\usepackage{float}

\usepackage{tikz}
\usetikzlibrary{matrix, positioning, fit, calc, arrows.meta, svg.path}
\usepackage{scalerel}

\usepackage{url}
\usepackage{comment}
\usepackage{lineno}
\usepackage[super]{nth}
\usepackage{datetime}
\usepackage{aligned-overset}

\usepackage{xr-hyper} 
\usepackage[colorlinks,citecolor=magenta,linkcolor=blue,filecolor=blue,urlcolor=blue]{hyperref}
\usepackage{cleveref}

\newtheorem{theorem}{Theorem}
\def\i{\mathrm{i}}
\newtheorem{lemma}{Lemma}

\theoremstyle{definition}
\newtheorem{remark}{Remark}
\newtheorem{definition}{Definition}
\newtheorem{example}{Example}
\newtheorem{mechanism}{Mechanism}

\definecolor{orcidlogocol}{HTML}{A6CE39}
\tikzset{
	orcidlogo/.pic={
		\fill[orcidlogocol] svg{M256,128c0,70.7-57.3,128-128,128C57.3,256,0,198.7,0,128C0,57.3,57.3,0,128,0C198.7,0,256,57.3,256,128z};
		\fill[white] svg{M86.3,186.2H70.9V79.1h15.4v48.4V186.2z}
		svg{M108.9,79.1h41.6c39.6,0,57,28.3,57,53.6c0,27.5-21.5,53.6-56.8,53.6h-41.8V79.1z M124.3,172.4h24.5c34.9,0,42.9-26.5,42.9-39.7c0-21.5-13.7-39.7-43.7-39.7h-23.7V172.4z}
		svg{M88.7,56.8c0,5.5-4.5,10.1-10.1,10.1c-5.6,0-10.1-4.6-10.1-10.1c0-5.6,4.5-10.1,10.1-10.1C84.2,46.7,88.7,51.3,88.7,56.8z};
	}
}
\newcommand\orcidicon[1]{\href{https://orcid.org/#1}{\mbox{\scalerel*{
				\begin{tikzpicture}[yscale=-1,transform shape]
					\pic{orcidlogo};
				\end{tikzpicture}
			}{|}}}}

\allowdisplaybreaks[4]
\graphicspath{{Pictures/}}

\makeatletter
\newcommand{\nket}[1]{\ket{#1^{\otimes N}}}

\newcommand{\Rmnum}[1]{\expandafter\@slowromancap\romannumeral #1@}
\newcommand{\sev}[1]{\expval{#1}_{S}}
\newcommand{\consc}[1]{\expval{#1}_{S,\text{connected}}}
\makeatother
\usepackage{times}

\usepackage{titlesec}
\titleformat{\section}[runin]
  {\itshape\normalsize\bfseries} 
  {\arabic{section}.} 
  {0.5em} 
  {}
\titlespacing*{\section}{0pt}{1ex}{0.5ex} 
\renewcommand{\thesection}{\arabic{section}}

\begin{document}

\title{Spurious Strange Correlators in Symmetry-Protected Topological Phases}

\author{Wei-Liang Gao}\thanks{These authors contributed equally.}
\affiliation{Guangdong Provincial Key Laboratory of Magnetoelectric Physics and Devices, State Key Laboratory of Optoelectronic Materials and Technologies,
	and School of Physics, Sun Yat-sen University, Guangzhou, 510275, China}

\author{Jie-Yu Zhang\orcidicon{0000-0003-4414-7680}}\thanks{These authors contributed equally.}\affiliation{Guangdong Provincial Key Laboratory of Magnetoelectric Physics and Devices, State Key Laboratory of Optoelectronic Materials and Technologies,
	and School of Physics, Sun Yat-sen University, Guangzhou, 510275, China}
\author{Zheng-Xin Liu}\email{liuzxphys@ruc.edu.cn}
\affiliation{School of Physics, Renmin University of China, Beijing, 100872, China}

\author{Peng Ye\orcidicon{0000-0002-6251-677X}}
\email{yepeng5@mail.sysu.edu.cn}
\affiliation{Guangdong Provincial Key Laboratory of Magnetoelectric Physics and Devices, State Key Laboratory of Optoelectronic Materials and Technologies,
	and School of Physics, Sun Yat-sen University, Guangzhou, 510275, China}

\begin{abstract}
Strange correlator is a powerful tool widely used in detecting symmetry-protected topological (SPT) phases.  However, the result of strange correlator crucially relies on the adoption of the reference state. In this work, we report that an ill-chosen reference state can induce spurious long-range strange correlators in trivial SPT phases, leading to false positives in SPT diagnosis. Focusing on 1D gapped bosonic/spin systems described by matrix product states (MPS), we trace the origin of these spurious signals in trivial SPT phases to the magnitude-degeneracy of the transfer matrix. We systematically classify three distinct mechanisms responsible for such degeneracy, each substantiated by concrete examples: (1) the presence of high-dimensional irreducible representations (abbreviated as \emph{irrep}) in the eigenspace corresponding to the entanglment spectrum (entanglement space); (2) a phase mismatch in symmetry representations between the target and reference states; and (3) long-range order arising from symmetry breaking. Our findings clarify the importance of the choice of proper reference states, providing a guideline to avoid pitfalls and correctly identify SPT order using strange correlators.
\end{abstract}

\maketitle

\section{Introduction------}
Symmetry-protected topological (SPT) phases in strongly interacting systems are short-range entangled (SRE) states that
cannot be adiabatically connected to a trivial product state without breaking a specific protecting symmetry~\cite{chen}. 
The Affleck--Kennedy--Lieb--Tasaki (AKLT) model serves as a paradigmatic example for understanding SPT phases~\cite{AKLT1}. Originally proposed as a solvable model for the spin-1 Haldane chain~\cite{Haldane}, its ground state can be precisely represented by a matrix product state (MPS)~\cite{MPS,MPSofAKLT2,MPSofAKLT3,book} constructed from a valence-bond-solid (VBS) picture~\cite{AKLT1,VBS1,VBS2,VBS3,highspin}. This framework has been instrumental in elucidating hallmark features of one-dimensional SPT phases such as string order~\cite{SO1,SO2,SO3,SO5}, symmetry-protected edge modes~\cite{AKLT1,book,Edge}, hidden symmetries~\cite{HS1,HS2,HS3,HS4,SO5}, and characteristic entanglement spectra~\cite{MPSofAKLT2,Pollmann}.

A key challenge in the field is to develop mathematical and physical tools that efficiently distinguish trivial from nontrivial SPT phases, thereby enabling a systematic classification and characterization of SPTs. A non-exhaustive set of representative approaches along this line includes those based on group cohomology~\cite{Chen2013CGLWbSPT}, cobordism groups~\cite{Kapustin2014Cobordism,Kapustin2015fCobordism}, nonlinear sigma models~\cite{Bi2015NLSMSPT,YouYou2016bSPT}, topological field theories~\cite{Lu2012CSSPT,bti2,Gu2016SPTBF,Ye20163dbSPT,2018arXiv180105416W}, conformal field theories~\cite{Hsieh2014SPT2dorbifold,Hsieh2016SPT3dorbifold,Han2017BCFTSPT}, the decoration picture~\cite{Chen:2014aa}, projective or parton constructions~\cite{YW12,LL1263,Ye14b,YW13a,ye16a,Wangsenthil2015}, topological response and gauged theories~\cite{PhysRevLett.112.141602,Hung_Wen_gauge,PhysRevD.88.045013,bti6,lapa17,Wang2015GaugeGravitySPT,PhysRevB.99.205120}, and braiding statistics approaches~\cite{levin_gu_12,wang_levin1,2016arXiv161209298P,ye17b}. Other frameworks have explored structural features such as low-dimensional edge excitations~\cite{PhysRevB.84.235141}, interacting systems~\cite{Chen_science}, and symmetry enrichment~\cite{ye16_set}.
Among these, the \textit{strange correlator}, originally proposed by You et al.~\cite{Wave2014You}, has seen rapid development and widespread application~\cite{Wave2014You,Quantum2015Wu,Strange2014Wierschem,Detection2016Wierschem,Bona2016He,Topological2017Zhong,sc23,sc22,sc21,sc26,sc25}. Recent advances have extended this tool beyond standard SPTs to diagnose subsystem SPTs~\cite{sc24}, and even the higher-order topological phases generated by cellular automata using multi-point strange correlators~\cite{sc25}.

Defined as a correlation function between a target state $\ket{\Phi}$ and a chosen trivial reference state $\ket{\Omega}$, the strange correlator is given by:
\begin{equation}\label{strange correlators}
    C(i,j) := \expval{O_i^a O_j^b}_S := \frac{\bra{\Omega} O_i^a O_j^b \ket{\Phi}}{\bra{\Omega}\ket{\Phi}},
\end{equation}
where $O_i^a$ and $O_j^b$ are local operators. The subscript $S$ emphasizes the distinction from conventional correlators where the bra and ket states are identical.
The standard criterion posits that if $\ket{\Phi}$ belongs to a nontrivial SPT phase {protected by symmetry group $G$} (i.e., possessing stable edge modes~\cite{chen}), the strange correlator should exhibit long-range behavior (power-law decay or saturation to a constant). Conversely, if the target state is a trivial SPT phase, the strange correlator is expected to decay exponentially or vanish.

While the strange correlator has achieved significant success, the rigorous criteria for constructing a valid setup—specifically the selection of the trivial reference state $\ket{\Omega}$—have not been systematically addressed. This ambiguity is critical because the conventional converse logic does not hold universally. In this work, we report the existence of \textit{spurious strange correlators}: long-range correlations that emerge even when the target state is topologically trivial.
A simple yet illustrating counterexample is the case of spin-1/2 trivial product states. Let $\ket{0},\ket{1}$ be the computational basis of a single spin-1/2 Hilbert space and $\ket{+}=\frac1{\sqrt 2}(\ket0 +\ket 1)$, then two product states
\begin{equation}\label{tps}
    \frac{\bra{0\cdots 0}X_iX_j\ket{+\cdots +}}{\bra{0\cdots 0}\ket{+\cdots +}}=1
\end{equation}
explicitly exhibit a long-range strange correlator, where $X_i:=\sigma^x_i$ is the Pauli-X operator acting on $i$-th spin. Here, the ratio ${\bra{\Omega}X_iX_j\ket{\Phi}}/{\bra{\Omega}\ket{\Phi}}$ equals unity regardless of distance\footnote{Strictly speaking, both the numerator and denominator vanish in the thermodynamic limit. As clarified in Ref.~\cite{Wave2014You}, calculations are based on finite system sizes before taking the thermodynamic limit of the ratio, which remains finite.}, exhibiting long-range behavior despite the absence of any SPT order.

This suggests that without a definitive selection rule for reference states, strange correlators may yield false positives. To the best of our knowledge, the mechanisms underlying such spurious behaviors and the guidelines for avoiding them have not been systematically explored. In this manuscript, we address this issue by providing a comprehensive analysis of \textit{strange correlators in 1D gapped bosonic/spin systems}.
By developing a matrix product state (MPS) framework, we prove theorems establishing that the long-range behavior of the  strange correlator can be attributed to the \emph{magnitude-degeneracy} of the largest eigenvalue of the transfer matrix (Sec.~\ref{theorems}).
Using this framework, we demonstrate that while nontrivial SPT phases necessarily exhibit this degeneracy due to Schur's lemma, they are not the sole source.
We systematically classify three distinct mechanisms responsible for spurious long-range correlations, each substantiated by concrete lattice models (Sec.~\ref{spurious1}):
(1) trivial phases involving high-dimensional irreducible representations, exemplified by the spin-2 AKLT model (\textit{Example} and \textit{Mechanism} \ref{HDirrep}) ;
(2) a phase mismatch between the symmetry representations of the reference and target states (\textit{Example} and \textit{Mechanism} \ref{phase-mismatch}); and
(3) long-range order arising from symmetry-breaking phases (\textit{Example} and \textit{Mechanism} \ref{symmetry-breaking}).

Our work provides a rigorous caveat for the application of strange correlators and offers analytic tools to unambiguously diagnose SPT phases by correctly selecting reference states. To preclude spurious signals from \textit{Mechanism \ref{phase-mismatch} and \ref{symmetry-breaking}}, the trivial reference state $|\Omega\rangle$ should be carefully chosen: it must (i) transform under the same 1D symmetry representation as the target state (preventing phase mismatch), and (ii) preserve the full global symmetry (distinguishing SPT order from symmetry-breaking long-range order). The spurious signals form \textit{Mechanism \ref{HDirrep}} are more intricate, and a more detailed operational guideline is presented in \textit{Remark \ref{remark2}}.

We focus our discussion on 1D systems with global symmetries; extensions to more exotic subsystem symmetries~\cite{sc25} are beyond the scope of this work.

\section{Origin of long-range strange correlators------}\label{theorems}

In this section, we aim to establish the analytical nature of the long-range strange correlator. We first write all correlators in the form of matrix product states (MPS), then prove that the long-range behavior of the connected strange correlator can be attributed to what we name as magnitude-degeneracy (explicit definition will be discussed below). 

We introduce the \textit{connected strange correlator}
\begin{eqnarray}
    \consc{O^a_iO^b_j}:=\sev{O^a_iO^b_j}-\sev{O^a_i}\sev{O^b_j}
\end{eqnarray}
 to reflect the long-range behavior of the strange correlator, where $\sev{O}:=\bra{\Omega}O\ket{\Phi}/\bra{\Omega}\ket{\Phi}$ is named strange expectation value. There are two reasons to do so:

 First, the existence of a long-range connected strange correlator guarantees the existence of operators yielding a long-range ordinary strange correlator (see Supplemental Material (SM) Sec.~\ref{stricter} for proof for \textit{Theorem \ref{theorem3}}), while the reverse is not true (Eq.~\eqref{tps} is a typical example).
Formally, we prove that
\begin{theorem}\label{theorem3}

Given a Matrix Product State (MPS) target state $|\Phi\rangle$ and a reference state $|\Omega\rangle$. If there exist local operators $O^a$ and $O^b$ such that the connected strange correlator exhibits long-range behavior in the thermodynamic limit, i.e.,
\begin{equation}
    \lim_{|i-j|\to\infty} \langle O^a_i O^b_j \rangle_{S, \text{connected}} \neq 0,
\end{equation}
then there exist (in fact, almost all) operators $\tilde{O}^a$ and $\tilde{O}^b$  such that the ordinary strange correlator also exhibits long-range behavior:
\begin{equation}
    \lim_{|i-j|\to\infty} \langle \tilde{O}^a_i \tilde{O}^b_j \rangle_{S} \neq 0.
\end{equation}
\end{theorem}

 Second, it is the connected part of strange correlators that relates to the contribution from the nontrivial SPT states, as shown in SM Sec.~\ref{nontrivial-SPT}. In fact, many cases discussed in the literature (e.g. AKLT models where $O^a_iO^b_j=S^+_iS^-_j$ and $\ket{\Omega}=\ket{0\cdots 0}$, 1D cluster states with $O^a_iO^b_j=\sigma^z_i\sigma^z_j$, etc.), these two correlators share the same value since $\sev{O^a}=\sev{O^b}=0$ in these cases.

Now we introduce the matrix product state (MPS) formalism to simplify the discussion, which provides a concise representation of 1D gapped states with arbitrary spins, and later we will see that the long-range behavior of the strange correlators will be reflected in the transfer matrix in MPS formalism.

Suppose the translational-invariant target state defined on $L$ spin-$S$ systems is represented by MPS:
\begin{equation}
\ket{\Phi}=\sum_{i_1,\cdots,i_L}\Tr\left[M^{i_1}\cdots M^{i_L} \right] \ket{i_1,\cdots,i_L},
\end{equation}
where each $i_n=-S,-S+1,\cdots,S-1,S$ reflects the physical dimension of the spin at each site. Here we have assumed the translational invariance and the periodic boundary condition.

\begin{figure}[htbp]
	\centering
\begin{tikzpicture}[scale=0.8]
\draw[thick] (0,0) -- (10,0);

\foreach \x in {1,2,3,4,6,7,8,9} {
    \fill[white] (\x,0) circle (0.4);
    \draw[thick] (\x,0) circle (0.4);
    \node at (\x,0) {$M$};
}

\fill[white] (3,1.1) circle (0.4);
\draw[thick] (3,1.1) circle (0.4);
\node at (3,1.1) {$O^a$};

\fill[white] (7,1.1) circle (0.4);
\draw[thick] (7,1.1) circle (0.4);
\node at (7,1.1) {$O^b$};

\foreach \x in {1,2,3,4,6,8,9,7} {
    \fill[black] (\x,2.2) circle (0.08);
    \node[above] at (\x,2.3) {$\vert\omega\rangle$};
}

\foreach \x in {1,2,4,6,8,9} {
    \draw[thick] (\x,2.2) -- (\x,0.4);
}

\draw[thick] (3,2.2) -- (3,1.5);
\draw[thick] (7,2.2) -- (7,1.5);

\foreach \x in {2.5,6.5,8.5} {
    \draw[thick] (\x,-0.5)--(\x,3)--(\x+1,3)--(\x+1,-0.5)--cycle;
}

\node at (3.05,-1.3) {$M_a$};
\node at (7.05,-1.3) {$M_b$};
\node at (9.05,-1.3) {$M_\omega$};

\draw[thick] (3,0.7) -- (3,0.4);
\draw[thick] (7,0.7) -- (7,0.4);

\node at (5,1.1) {\large$\cdots$};
\node at (0.5,1.1) {\large$\cdots$};
\node at (10,1.1) {\large$\cdots$};
\end{tikzpicture}
    \caption{The formulation of $\expval{O^aO^b}_S$ articulated utilizing a Matrix Product State (MPS) representation. The tensor denoted as $M$, which is depicted with three interconnected lines, corresponds to the $M$ tensor within the MPS framework, characterized by two horizontal virtual legs and a single vertical physical leg. The black dots are indicative of direct product states, each defined by an individual physical index. The interconnecting lines signify the process of index contraction. The definitions of $M_a$, $M_b$, and $M_\omega$ are each enclosed within a rectangular boundary.}
    \label{fig3:image1}
\end{figure}
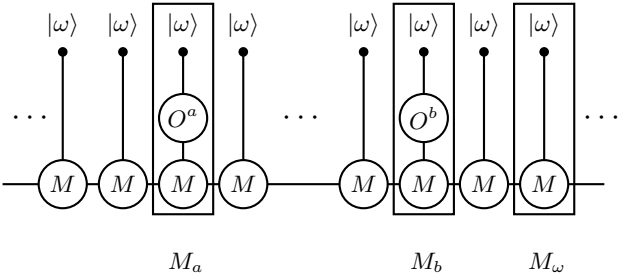

The calculation of the strange correlator in MPS form reduces to calculating the trace of certain matrices:
\begin{equation}\label{sc_def}\begin{aligned}
		\expval{O^a_iO^b_{i+R+1}}_S = \frac{\Tr\left[M_a M_\omega^R M_b M_\omega^{L-R-2} \right]}{\Tr\left[M_\omega^L \right]}.
\end{aligned}\end{equation}
where $L$ is the length of the system and $R$ is the distance between two operators. We denote the reference state as $|\Omega\rangle=\ket{\omega}\otimes...\otimes \ket{\omega}$, and here
$M_a:= \sum_\sigma \bra{\omega}O^a\ket{\sigma} M^{\sigma},~M_b:= \sum_\sigma \bra{\omega}O^b\ket{\sigma} M^{\sigma},~M_\omega:= \sum_\sigma \bra{\omega}\ket{\sigma} M^{\sigma}.$
We refer to the matrices $ M_a $ and $ M_b $ as \textit{operator matrices}, and $ M_\omega $ as the \textit{transfer matrix}. These matrices are pictorially shown in Fig.~\ref{fig3:image1}. Similarly, we can calculate the strange expectation value (SEV):
\begin{equation}\label{sev_def}\begin{aligned}
		\expval{O^a_i}_S =\frac{\Tr\left[M_a  M_\omega^{L-1} \right]}{\Tr\left[M_\omega^L \right]}.
\end{aligned}\end{equation}

In the MPS form developed above, 
we can easily observe that the long-range connected strange correlators originate from the \textit{magnitude-degeneracy} of the largest-magnitude eigenvalue of the transfer matrix. 
Here, we define:

\begin{definition}[Magnitude-degeneracy]
   An eigenvalue $\lambda_0$ of a matrix $M$ is said to be $D$-fold magnitude-degenerate iff there exist $\lambda_1,\lambda_2,\cdots,\lambda_{D-1}$ such that they share the same magnitude, i.e.
        $|\lambda_0|=|\lambda_1|=|\lambda_2|=\cdots |\lambda|_{D-1}.$
\end{definition}

Formally, we prove that:
\begin{theorem}[Origin of long-range strange correlator]\label{magnitude-deg}
Suppose the eigenvalue of $M_\omega$ with the largest magnitude is $D$-fold magnitude-degenerate. 
i.e. we assume that
   $ |\lambda_0|=|\lambda_1|=\cdots=|\lambda_{D-1}|>|\lambda_{D}|>\cdots.$

If $D=1$, then $\forall O^a,O^b$, the connected strange correlators decay exponentially to zero in the long-distance limit:
\begin{equation}\label{csc}
    \expval{O^aO^b}_{S,\text{connected}}:=\expval{O^aO^b}_S-\expval{O^a}_S\expval{O^b}_S=0.
\end{equation}

    If $D>1$, then there exist some operator $O^a,O^b$  and an infinite sequence $\{R_i\}:=R_1,R_2,\cdots$ where $j>i\iff R_j>R_i>0$ and $\lim_{i\to\infty} R_i=\infty$, such that
\begin{equation}
    \lim_{k\to \infty} \expval{O^a_iO^b_{i+R_k+1}}_{S,\text{connected}}\ne 0.
\end{equation}
\end{theorem}

We see in \textit{Theorem \ref{magnitude-deg}} that, similar to the behavior of normal correlation functions in MPS formalism, the behavior of strange correlators also significantly depends on the eigenvalue with the largest magnitude. 

The complete proof of the theorem is presented in SM Sec.~\ref{tm1}.

Finally, one may worry about whether it is hard to numerically search $O^a,O^b$ such that strange correlators exhibit long-range behavior when the largest-magnitude eigenvalue is magnitude-degenerate, as described in \textit{Theorem \ref{magnitude-deg}}. We claim that almost every local operator satisfies the requirement above.  Formally, we prove that
\begin{theorem}\label{theorem2}
    If $D>1$, we define the direct sum of spaces of operator matrices $M_a, M_b$ as $\mathcal M:=\{(M_a,M_b)|M_a := \sum_\sigma \bra{\omega}O^a\ket{\sigma} M^{\sigma} ,
		M_b := \sum_\sigma \bra{\omega}O^b\ket{\sigma} M^{\sigma}\}$, then $\expval{O^a_iO^b_j}_{S,\text{connected}}$ does not converge to zero (i.e. exhibits long-range behavior) almost everywhere in $\mathcal M$, where the exceptions have zero Lebesgue measure in $\mathcal M$.
\end{theorem}

Roughly speaking, this is proved by the fact that the measure of the zero point set of a not identically zero analytic function is zero (\textit{Lemma~\ref{zero-point}} in SM), of which the details are presented in SM Sec.~\ref{tm2}.

For a SPT state, the virtual bond in the MPS representation carries projective representation (abbreviated as \textit{rep}) of the symmetry group $G$ \cite{chen}. Namely, suppose the symmetry acts uniformly on the system as $u(g)\otimes u(g)\cdots$ on each site, where $u(g)$ is a unitary representation of $g\in G$ in the local Hilbert space, then 
\begin{equation}\label{ugM}
    \sum_{j} [u(g)]_{ij} M^{j} = \alpha(g) V^{-1}(g) M^{i} V(g),
\end{equation}
where $V(g)$ is a projective rep of $G$ characterized by the invariant $\beta(G)$ being the factor system, and $\alpha(g)$ is a 1-dimensional (1D) linear rep of $G$. If the reference state $|\Omega\rangle=\ket{\omega}\otimes...\otimes \ket{\omega}$ is a symmetric trivial state with $|\omega\rangle =\sum_i \omega_i|i\rangle,~u(g)|\omega\rangle =\sum_{i,j} u_{ij}\omega_j|i\rangle= e^{i\theta(g)}|\omega\rangle$, then the the transfer matrix $M_\omega$ vary as
\begin{equation}\label{VMV}
 V(g) M_\omega V^{-1}(g)  =   A(g) M_\omega,
\end{equation}
with $A(g)=\alpha(g)e^{-i\theta(g)}$ a 1D linear rep of $G$. For finite groups (also for some compact Lie groups), there exist a finite integer $N$ such that $[A(g)]^N=1$ for all $g\in G$, then one has $ V(g) [M_\omega]^N V^{-1}(g)  =  [M_\omega]^N$.

It can be shown that when the projective rep $V(g)$ belongs to a nontrivial class,  the eigenvalues of $M_\omega$ are magnitude degenerate (SM Sec.~\ref{nontrivial-SPT}), giving rise to nonzero strange correlator. 

\section{Spurious Strange Correlators: Examples and Mechanisms------}\label{spurious1}

{However, the magnitude-degeneracy of eigenvalues can also appear in a trivial SPT phase,}
and so does the long-range strange correlator. This mechanism shares the same mathematical origin (Schur's lemma) as nontrivial SPT phases (SM Sec.~\ref{nontrivial-SPT}), but arises in a topologically trivial phase.

\begin{example}[$SO(3)$ $D=3$ irrep]\label{so3irrep}
    Consider  the ground state of spin-2 AKLT model. This state has $SO(3)$ symmetry and belongs to the trivial SPT phase.
    The MPS form of spin-2 AKLT ground state is (see SM Sec.~\ref{mps} for MPS form of AKLT states):
\begin{equation}
\begin{aligned}
&M_{-2}=\smqty(0&0&\sqrt{3/5}\\0&0&0\\0&0&0), 
M_{-1}=\smqty(0&-\sqrt{3/10}&0\\0&0&\sqrt{3/10}\\0&0&0),\\
&M_0=\smqty(\sqrt{1/10}&0&0\\0&-\sqrt{2/5}&0\\0&0&\sqrt{1/10}),\\
&M_1=\smqty(0&0&0\\\sqrt{3/10}&0&0\\0&-\sqrt{3/10}&0), 
M_2=\smqty(0&0&0\\0&0&0\\\sqrt{3/5}&0&0).
\end{aligned}
\end{equation}
Under $SO(3)$, 
the matrix $M_i$ transforms as (\ref{ugM}), where $V(g)$ is a $D=3$ linear irrep of $SO(3)$ and $\alpha(g)=1$.

    Since the system possesses $SO(3)$ invariance, we require the reference state to also exhibit $SO(3)$ invariance. However, the Hilbert space of a single spin cannot support such a spin singlet. Yet, the direct sum decomposition of the tensor product of two spins contains a spin singlet component. Therefore, we consider two adjacent spins together, yielding the following $SO(3)$-invariant state:
    $\ket{\omega}=\frac{1}{\sqrt 5}(\ket{2,-2}-\ket{1,-1}+\ket{0,0}-\ket{-1,1}+\ket{-2,2}),$
    thus $A(g)=\alpha(g)e^{-\i\theta(g)}=1$ in this case. Note that we have merged two sites into a new site to obtain a $SO(3)$ invariant reference state.

    Here both target state and reference state are $SO(3)$ symmetric. Explicitly calculation of the transfer matrix yields
 $M_\omega=\frac{1}{\sqrt{5}}\mathbb I_{3\times 3}  $
    explicitly showing a 3-fold magnitude-degeneracy. Here we pick $O^a=S^+\otimes I,~O^b=I\otimes S^-$ with $I$ being the identity, then 
$\sev{O^a}=\sev{O^b}=0$ and
    \begin{eqnarray}
        \consc{O^aO^b}=\sev{O^aO^b}=-3,
    \end{eqnarray}
    explicitly showing spurious strange correlator.
\end{example}

\begin{mechanism}[$D>1$ irrep]\label{HDirrep}
{If the MPS is a trivial SPT with $\beta=1$, then in this case, the dimensionality of the irreducible linear reps of $G$ can be either 1 or greater than 1. All irreps of Abelian groups are 1-dimensional, while non-Abelian groups can host higher dimensional irreps. In this case, high order magnitude degeneracy is not guaranteed in general. However, if the eigenspace of the maximal eigenvalue of $(M_\omega)^N$ is a high dimensional irrep by coincidence, then the magnitude-degeneracy may still be realized. }

\end{mechanism}

It is practical to clarify when we can avoid spurious signals from \textit{Mechanism \ref{HDirrep}}, and how to numerically choose reference states. We provide a concise procedure below.

\begin{remark}[Operational Guideline for Removing Spurious Signals from \textit{Mechanism \ref{HDirrep}}] \label{remark2}

We propose an algebraic protocol for numerical implementations for choosing reference state that avoids \textit{Mechanism \ref{HDirrep}}. 
Given a target MPS in trivial phase with virtual bond possessing high-dimensional  irreps, one can first decompose its virtual space $\mathbb{V}$ (the representation space of $V(g)$) under the symmetry group $G$ into irreducible sectors: $\mathbb{V} \cong \mathbb{V}_{\text{triv}} \oplus \mathbb{V}_{\text{high}}$, where $\mathbb{V}_{\text{triv}}$ denotes the 1D trivial sector (signal) and $\mathbb{V}_{\text{high}}$ collects all high-dimensional representations (noise). 
The validity of a reference state $|\Omega\rangle$ relies on the spectral gap of the transfer matrix: $|\lambda_{\text{triv}}| > |\lambda_{\text{high}}|$.
An optimal reference state vector $\vec{\omega}$ can be constructed deterministically by maximizing the signal-to-noise ratio (SNR). 
Define the signal vector $\vec{t} = \text{Tr}_{\mathbb{V}_{\text{triv}}}[M]$ and the noise covariance matrix $\mathbf{K} = \text{Tr}_{\mathbb{V}_{\text{high}}}[M \otimes M^\dagger]$ acting on the physical space, where $\Tr_{\mathbb{V}_{\text{triv/high}}}[M]\equiv \Tr[P_{\text{triv/high}}M]$ with $P_{\text{triv/high}}$ being the projector onto the corresponding space. The optimal reference is given by the generalized inverse solution $\vec{\omega}_{\text{opt}} \propto \mathbf{K}^{+} \vec{t}$ \cite{Penrose_1955}, where $\mathbf{K}^{+}$ is the Moore-Penrose inverse of $\mathbf{K}$ . 
If the single-site $\vec{\omega}_{\text{opt}}$ is insufficient to let $|\lambda_{\text{triv}}| > |\lambda_{\text{high}}|$, one can perform coarse-graining (blocking two or more sites) to enlarge the physical Hilbert space. 

We emphasize that the success of this algorithm relies on the existence of at least one one-dimensional irrep in the entanglement space. In the case of nontrivial SPT phases, where the spectrum of transfer matrix $M_\omega$ is enforced to be magnitude-degenerate due to projective representations, such a unique scalar channel is fundamentally absent.
\end{remark}

Now we show another mechanism that also leads to magnitude-degeneracy. We provide a clean example below devoid of any higher-dimensional irreps of $V(g)$, thus \textit{Mechanism~\ref{HDirrep}} is not active. First we present following example:

\begin{example}[$\mathbb Z_2$ phase mismatch]
Consider an MPS with physical spin and virtual spin are both 1/2, which leads to a $D=2$ bond dimension.  Now we suppose the MPS has a global $\mathbb Z_2$ spin flip symmetry (uniform action of Pauli-$X$ operator). This choice excludes the effect of the previously discussed higher-dimensional irreps of $V(g)$, since $H^2(\mathbb Z_2,U(1))=1$ and that $\mathbb Z_2$ is Abelian. This means that $\mathbb Z_2$ symmetry cannot yield a nontrivial factor system, and cannot have a higher-dimensional irrep because of its Abelian nature. 

This symmetry constraint can be explicitly expressed as 
    $XM_0=\alpha(g)ZM_0Z,~XM_1=\alpha(g)ZM_1Z$
where $X,Z$ are Pauli-$X,Z$ operators, and $\alpha(g)=\pm1$ with $g$ denoting the spin flip action. 

The general form of MPS satisfying such condition can be written as
\begin{equation}
    M_0=\begin{pmatrix}
        u&v\\w&x
    \end{pmatrix},~
    M_1=\alpha(g)\begin{pmatrix}
        u&-v\\-w&x
    \end{pmatrix}
\end{equation}
where $u,v,w,x\in\mathbb C.$

Such generally written MPS is always injective except several fine-tuned points, so we can always find a frustration-free parent Hamiltonian whose unique gapped ground state is the chosen MPS \cite{Fannes1992}.
We choose the reference state as
    $\ket{\Omega}=\bigotimes_i \ket{\omega},$
where $\ket{\omega}=\omega_0\ket{0}+\omega_1\ket 1.$

Define $\omega_\pm^*=\omega_0^*\pm\alpha(g)\omega_1^*$ with $*$ symbol denoting the complex conjugate, 
then the transfer matrix is 
\begin{eqnarray}
    M_\omega=\begin{pmatrix}
        u\omega_+^*&v\omega_-^*\\w\omega_-^*&x\omega_+^*
    \end{pmatrix}.
\end{eqnarray}
Here,
$\lambda_{\pm}=\frac12{(u+x)\omega_+^*\pm \frac12[{(u-x)\omega_+^{*2}+4vw\omega_-^{*2}}}]^{1/2}$ is the eigenvalues of the tranfer matrix.
It is obvious that if we choose
 $ \omega_+^*=0$, which means $\omega_0=-\alpha(g) \omega_1 $ and $X\ket{\omega}=-\alpha(g)\ket{\omega}$, then it follows that the eigenvalues 
$
     \lambda_{\pm}={\pm\omega_-^*\sqrt{vw}},$
explicitly demonstrating the magnitude-degeneracy regardless of the choice of $u,v,w,x$ (assuming $vw\ne0$), thus causing a long-range connected strange correlator. 
 
 If we pick $O^a_i=Z_i,O^a_j=Z_j,\alpha(g)=1,\ket{\omega}=1/\sqrt 2 (\ket0-\ket 1)$, then in the thermodynamic limit (system length $L$ is set to be even while taking the limit to keep the correlator well-defined), the strange correlators reads
 \begin{equation}
    \sev{Z_iZ_{i+R}}=\consc{Z_iZ_{i+R}}=\begin{cases}
        \frac{u^2+x^2}{2vw},&R \text{ is odd;}\\
        \frac{ux}{vw},& R \text{ is even}.
    \end{cases}
 \end{equation}
following the definition of Eq.~\eqref{sc_def}, \eqref{sev_def}, \eqref{csc}.
 
For randomly selected $u,v,w,x$, such correlator is almost always nonzero.

\end{example}

The example above can be summarized as the target state and the reference have the same symmetry, but produce different phases under the symmetry action. We also show an example with parity symmetry.

\begin{mechanism}[Phase Mismatch]\label{phase-mismatch}
Suppose the MPS target state has an onsite symmetry group $G$.

Supposing that the product rep $V(g)\otimes [V^{-1}(g)]^T$ contains the 1D rep $A(g)=\alpha(g)e^{-i\theta(g)}$, and the reference state $\ket{\Omega}$ satisfy $g\ket{\omega}=e^{\i\theta(g)} \ket{\omega}$, or equivalently $\sum_j u(g)_{ij}\omega_j=e^{\i\theta(g)}R(g)^{-1}\omega_iR(g) =e^{\i\theta(g)}\omega_i$ with $R(g)$ a 1D representation of $G$, then one has  Eq.~(\ref{VMV}), namely $ M_\omega  V^{-1}(g) = A(g)V^{-1}(g) M_\omega$. If $A(g)\neq 1$ for some $g\in G$, then the transfer matrix can be magnitude-degenerate even without the higher dimensional irrep in $V(g)$. For any eigenvector $M_\omega \ket{\lambda}=\lambda\ket{\lambda}$, $V(g)^{-1}\ket{\lambda}$ is another eigenvector with the eigenvalue  $\lambda A(g)$ having the same magnitude:
  $  M_\omega V(g)^{-1}\ket{\lambda} = \lambda A(g) V(g)^{-1}\ket{\lambda}.$
    
    This argument can be also applied to the case of parity symmetry, where the parity transformation $P=w\otimes w\otimes\cdots \otimes w P_1$ with $w$ being the onsite unitary transformation satisfying $w^2=1$ and $P_1$ being the exchange of sites. See \textit{Example~\ref{parity-mismatch}} in SM Sec.~\ref{more-examples}. There are also cases where different mechanism can act simultaneously. \textit{Example~\ref{mixed-example}} in SM Sec.~\ref{more-examples} shows the case where  \textit{Mechanism~\ref{HDirrep}} and \textit{Mechanism~\ref{phase-mismatch}} act simultaneously.

\end{mechanism}

Two mechanisms above have assumed that the MPS target state is the unique symmetric ground state of the system Hamiltonian, i.e. no symmetry is broken. Now we examine the case where the symmetry is broken.

\begin{example}[GHZ state]\label{ghz}
    As a concrete example, we consider the 1D Ising model with nearest neighbor $ZZ$ coupling. We choose $\ket{\Phi}=\frac1{\sqrt 2}\left(\nket{0}+\nket{1}\right)$ and  $\ket{\Omega}=\nket{+}$, and $O^a_iO^b_j=Z_iZ_j$. 
Now we check explicitly that
\begin{equation}
    M_0=\begin{pmatrix}
        1&0\\0&0
    \end{pmatrix},~  M_1=\begin{pmatrix}
        0&0\\0&1
    \end{pmatrix},
\end{equation}
thus
    $M_\omega=\operatorname{diag}(2^{-1/2},2^{-1/2}),$
exhibiting magnitude-degeneracy. 

It follows that
\begin{equation}
    \begin{aligned}
        \expval{Z_iZ_j}_S&=\frac{\bra{+^{\otimes N}}Z_iZ_j\ket{\Phi}}{\bra{+^{\otimes N}}\ket{\Phi}}=\frac{2^{-N/2}+2^{-N/2}}{2^{-N/2}+2^{-N/2}}=1,
    \end{aligned}
\end{equation}
and
\begin{equation}
    \begin{aligned}
        \expval{Z_i}_S&=\frac{\bra{+^{\otimes N}}Z_i\ket{\Phi}}{\bra{+^{\otimes N}}\ket{\Phi}}=\frac{2^{-N/2}-2^{-N/2}}{2^{-N/2}+2^{-N/2}}=0,
    \end{aligned}
\end{equation}
exhibiting the long-range strange correlator:
\begin{eqnarray}
    \consc{Z_iZ_j}=\sev{Z_iZ_j}=1.
\end{eqnarray}

While a GHZ state is not classified as a nontrivial SPT state, it is indeed highly entangled. We see explicitly here that strange correlators also give a long-range response to such entanglement structure.
\end{example}

\begin{mechanism}[Symmetry breaking]\label{symmetry-breaking}

In previous sections, we have assumed the MPS target state is the unique symmetric ground state of the system. In fact, in the symmetry breaking case, there are also mechanisms that lead to the magnitude-degeneracy of the transfer matrix, thus the long-range strange correlator.

    The origin of the long-range strange correlators can be explained as follows. If the system is in a symmetry breaking phase, the unique symmetric ground state of the system (some linear combinations of ground state basis, e.g. GHZ state $\ket{0\cdots 0}+\ket{1\cdots 1}$ is a symmetric ground state of 1D Ising model with $H=\sum_{i}Z_iZ_{i+1}$, preserving the global spin flip symmetry) will have a block-diagonalized MPS form \cite{chen}
  $
        M_i=\operatorname{diag}(M^{(0)}_i,M^{(1)}_i,\cdots)$
    with number of blocks greater than 1 and the double tensor $\mathbb E^{(k)}=\sum_{i}M_i^{(k)}\otimes M_i^{(k)*}$ for each block sharing the same largest eigenvalue. We claim that one can always pick a symmetric reference state to make the transfer matrix to be magnitude-degenerate, where
       $ M_\omega=M_i\omega_i^*=\operatorname{diag}(M^{(0)}_i\omega_i^*,M^{(1)}_i\omega_i^*,\cdots),$
as proved in SM Sec.~\ref{ssb-proof}. In \textit{Example~\ref{ghz}}, the reference state is chosen as $\ket{+\cdots +}$, being symmetric under the spin flip symmetry (global Pauli-$X$ action). 
    
    In this scenario, the magnitude-degeneracy come from different blocks from the MPS tensor, rather than the high dimensional irreps within a single block. The mechanisms discussed above can exist simultaneously and only by careful selection of the reference state can one use strange correlators to correctly detect nontrivial SPT phases.
\end{mechanism}

\section{Discussion and Outlook------} \label{sec:discussion}

The strange correlator is a widely used probe for SPT order, but its outcome depends crucially on the choice of reference state. Within the MPS framework, we show that a long-range strange correlator stems directly from a magnitude-degeneracy in the transfer matrix's leading eigenvalues. To avoid false positives, a valid reference state must not only be trivial, but also: (i) preserve the full global symmetry, (ii) match the 1D symmetry representation of the target state (e.g., carry the same parity), and (iii) be chosen to ensure that high-dimensional irreps in the entanglement space do not lead to the dominant eigenvalues of the transfer matrix. 

This finding provides a practical guide for numerical studies like quantum Monte-Carlo (QMC) or exact diagonalization (ED), where the ground state is not known in MPS form. For these cases, we propose an ``inverse scanning" strategy. The key idea is that the strange correlator probes the relative SPT invariant between the target and reference states. For a true SPT phase, the virtual space carries a projective representation (like a half-integer spin), a property that cannot be lifted by any reference state transforming in a linear representation (like an integer spin). The resulting eigenvalue degeneracy of the transfer matrix is therefore robust.
In contrast, the spurious signal in a trivial phase (like the spin-2 AKLT model) stems from high-dimensional linear representations that are not topologically protected and are therefore fragile. In a trivial phase, generic symmetric perturbations will mix the scalar representation (a 1D irrep) into the virtual space. This effect can be captured by a well-chosen reference state (see \textit{Remark \ref{remark2}}). In practice, one can test the target state against a family of symmetric reference states, perhaps using coarse-graining (blocking sites) to explore a larger space of trivial states. If the long-range correlation is robust across the entire family of references, it is a strong indicator of genuine SPT order. Conversely, if the signal is fragile — meaning a specific reference state can be found that results in short-range decay — the target state is identified as trivial.

We also comment that the spurious signal in \textit{Mechanism \ref{HDirrep}} has a direct physical interpretation in terms of ``representation SPT" (RSPT) phases \cite{OBrien2020}.  An RSPT phase has an MPS whose virtual bonds carry a high-dimensional linear irrep of the symmetry group (e.g., the 3D irrep of SO(3) in \textit{Example \ref{so3irrep}}). This makes it distinct from a simple product state, even though they belong to the same trivial group cohomology class. A deeper study of this connection is left for future work.

{We can also relate these findings to the entanglement spectrum (ES). For a nontrivial SPT phase, the degeneracies in both the ES and the strange correlator are protected. For a trivial phase with a non-Abelian symmetry, the ES can still be degenerate, but this degeneracy can be lifted by a boundary phase transition without closing the bulk gap. The same is true for the strange correlator. However, a non-vanishing strange correlator does not necessarily imply a ES degeneracy of the maximal entanglement weight. For instance, in \textit{Mechanism \ref{phase-mismatch}}, a trivial phase with an Abelian symmetry can have a non-degenerate ES, but still produce a long-range strange correlator if the reference state is chosen poorly.
}

Finally, our rigorous results are based on the exact contractibility of 1D bosonic MPS. Extending this analysis to higher dimensions (e.g., using PEPS) requires care, as approximate contraction schemes can obscure the clean spectral properties of the transfer matrix. For fermionic systems, one must use graded tensor networks and ensure the reference state correctly matches the target's fermion parity structure to avoid spurious cancellations. These extensions present challenges for future work. 
\acknowledgements
This work was in part supported by National Natural Science Foundation of China (NSFC)  under Grants No. 12474149, Research Center for Magnetoelectric Physics of Guangdong Province under Grant No. 2024B0303390001, and Guangdong Provincial Key Laboratory of Magnetoelectric Physics and Devices under Grant No. 2022B1212010008. ZXL acknowledge the support from NSFC 12374166, NSFC 12134020 and National Key Research and Development Program of China (Grant No. 2022YFA1405300, 2023YFA1406500).

%

\clearpage
\onecolumngrid
\appendix
    
\setlength{\baselineskip}{1.5\baselineskip}
\titleformat{\section}
  {\large\bfseries\MakeUppercase} 
  {\Alph{section}.} 
  {1.5em} 
  {}
\titlespacing*{\section}{0pt}{1ex}{0.5ex} 

\renewcommand{\thesection}{\Alph{section}}

\title{Supplemental Material for ``Spurious Strange Correlators in Symmetry-Protected Topological Phases''}
\author{Wei-Liang Gao}\thanks{These authors contributed equally.}
\affiliation{Guangdong Provincial Key Laboratory of Magnetoelectric Physics and Devices, State Key Laboratory of Optoelectronic Materials and Technologies,
	and School of Physics, Sun Yat-sen University, Guangzhou, 510275, China}

\author{Jie-Yu Zhang\orcidicon{0000-0003-4414-7680}}\thanks{These authors contributed equally.}\affiliation{Guangdong Provincial Key Laboratory of Magnetoelectric Physics and Devices, State Key Laboratory of Optoelectronic Materials and Technologies,
	and School of Physics, Sun Yat-sen University, Guangzhou, 510275, China}
\author{Zheng-Xin Liu}\email{liuzxphys@ruc.edu.cn}
\affiliation{School of Physics, Renmin University of China, Beijing, 100872, China}

\author{Peng Ye\orcidicon{0000-0002-6251-677X}}
\email{yepeng5@mail.sysu.edu.cn}
\affiliation{Guangdong Provincial Key Laboratory of Magnetoelectric Physics and Devices, State Key Laboratory of Optoelectronic Materials and Technologies,
	and School of Physics, Sun Yat-sen University, Guangzhou, 510275, China}

\maketitle
\section{Brief Review: Classification of Gapped Phases in 1D}\label{review}

In Ref.~\cite{chen}, a complete classification of 1D gapped phases is presented using the group cohomology theory. In the framework of Matrix Product States (MPS), these phases are characterized by the projective representations of the symmetry group on the virtual bonds.

Consider a translationally invariant MPS $|\Psi\rangle$ composed of tensors $M$. If the state is invariant under an onsite unitary symmetry $G$, the fundamental theorem of MPS implies that the action of the symmetry $u(g)$ on the physical indices induces a gauge transformation on the virtual indices:
\begin{equation}
    \sum_{j} [u(g)]_{ij} M^{j} = \alpha(g) V^{-1}(g) M^{i} V(g),
\end{equation}
where $g \in G$, $M^i$ are the MPS matrices, and $V(g)$ is a unitary matrix acting on the virtual bond. This equation introduces two crucial quantities that label the gapped phases:

\begin{enumerate}
    \item \textbf{The Projective Representation (Factor System) $\omega$:} The matrix $V(g)$ forms a projective representation of the group $G$. Since the physical symmetry is a linear representation ($u(g)u(h)=u(gh)$), the associativity of the gauge transformation requires $V(g)$ to satisfy:
    \begin{equation}
        V(g)V(h) = \omega(g,h) V(gh),
    \end{equation}
    where $\omega(g,h) \in U(1)$ is the factor system (or 2-cocycle). The equivalence classes of $\omega$ are classified by the second cohomology group $H^2(G, U(1))$. A nontrivial class $[\omega] \neq [1]$ indicates a nontrivial SPT phase protected by $G$, which is directly associated with gapless edge modes (degeneracy of the entanglement spectrum).

    \item \textbf{The 1D Linear Representation $\alpha(g)$:} The phase factor $\alpha(g)$ appearing in the MPS transformation rule must form a 1D linear representation of $G$ (for linear onsite symmetries) to ensure consistency. This implies $\alpha(g)\alpha(h) = \alpha(gh)$. Crucially, $\alpha(g)$ characterizes the charge or the quantum number of the state under symmetry $G$ but does \textit{not} protect gapless edge modes.
\end{enumerate}

The classification labels for common symmetries are summarized in Tab.~\ref{phase-labels}.

\begin{table}[h]
    \centering
    \renewcommand{\arraystretch}{1.3}
    \begin{tabular}{c|c|c}
       \hline \hline
       Symmetry &  ``Phase'' Label& SPT Label\\
       & (1D Rep / Charge) & (Factor System) \\
       \hline
       Onsite Unitary $G$ & $\alpha(g)$ & $\omega(g,h)$ \\
       Time Reversal $T$ & - & $\beta(T)$ \\
       Parity $P$ & $\alpha(P)$ & $\beta(P)$ \\
       \hline \hline
    \end{tabular}
    \caption{Classification of 1D bosonic gapped phases under various symmetries. The ``SPT Label'' (related to the factor system or projective representation) dictates the existence of edge modes. The ``Cohomology Label'' (related to 1D representations) distinguishes phases that are Short-Range Entangled (SRE) but do not necessarily possess SPT order.}
    \label{phase-labels}
\end{table}

To clarify the physical meaning of $\alpha(g)$, consider the global spin flip symmetry $\mathbb Z_2=\{e, a\}$ generated by the Pauli-$X$ operator. The group elements satisfy $a^2=e$. 
Consider two trivial product states: the ferromagnetic state $|\Psi_-\rangle = \ket{-\cdots -}$ (where $X\ket{-} = -\ket{-}$) and the state $|\Psi_+\rangle = \ket{+\cdots +}$ (where $X\ket{+} = +\ket{+}$).
\begin{itemize}
    \item For $|\Psi_-\rangle$, the site-local action gives a phase of $-1$. In MPS language, this corresponds to a 1D representation $\alpha(a) = -1$.
    \item For $|\Psi_+\rangle$, the site-local action gives a phase of $+1$, corresponding to $\alpha(a) = +1$.
\end{itemize}
While these two states cannot be connected by finite depth local unitary circuits (when the system length is odd), but these two states are usually recognized within the trivial phase. We see strange correlators also give response to difference of such phase labels.

In the context of strange correlators, we emphasize that only the quantities in the third column of Table~\ref{phase-labels} ($\omega, \beta(T), \beta(P)$) correspond to nontrivial SPT order. A nontrivial $\alpha(g)$ or $\alpha(P)$ merely indicates a specific 1D representation (e.g., a charge or parity eigenvalue) and does not inherently lead to the long-range entanglement signatures usually sought in strange correlators. However, as discussed in the main text (\textit{Mechanism \ref{phase-mismatch}}), a mismatch of this $\alpha$ label between the target and reference states can lead to spurious long-range correlations.

\section{Nontrivial SPT phase and Magnitude-Degeneracy}\label{nontrivial-SPT}
In the main text we have demonstrated that the magnitude-degeneracy of the transfer matrix will lead to long-range strange correlators. Now we show how a nontrivial SPT state contribute to the magnitude-degeneracy of the transfer matrix.

\begin{mechanism}[Nontrivial SPT phase]\label{SPT}
Consider the case of global onsite unitary symmetry $G$.
We assume that the original MPS is injective, namely, the maximum eigenvalue of the double tensor, namely, $E = \sum_i M_i^*\otimes M_i$ is nondegenerate. 

Consider the reference product state $|\Omega\rangle$ which is symmetric under any $u(g),~g\in G$, where $u(g)$ is a representation of $G$. That is to say, the reference state is $\ket{\Omega}=\bigotimes_i\ket{\omega}_i$ ,$\ket{\omega}:=\omega_i\ket{i}$, and $u(g)$ satisfies
      $  u(g)\ket{\omega}=e^{\i\theta(g)} \ket{\omega},$
    or equivalently, when the reference state is expressed in MPS form,
       $ u(g)_{ij}\omega_j=e^{\i\theta(g)}R^{-1}(g)\omega_iR(g)$
    where  and $R(g)$ being a representation of $G$. 

The MPS tensor transforms under $G$ via following rules \cite{chen}:
$   u(g)_{ij}M_j=\alpha(g)V^{-1}(g)M_iV(g),$
where
$\alpha(g)$ is a 1D rep of $G$, $V(g)$ is a projective rep (may be trivial) of $G$.
Here the transfer matrix is $M_\omega=M_i\omega_i^*$, with $*$ symbol denoting complex conjugate. Then we have
$\omega_iu(g)_{ij}M_j=\alpha(g)V^{-1}(g)M_i\omega_i^*V(g)$, or equivalently,
$  V(g)  M_i\omega_i^*= \alpha(g)e^{-\i\theta(g)} M_i\omega_i^*V(g).$
It follows that
\begin{eqnarray}\label{VMV}
V(g)  M_\omega  V^{-1}(g) = \alpha(g)e^{-\i\theta(g)}M_\omega,
\end{eqnarray}
Treating $M_\omega$ as a vector, then Eq.~(\ref{VMV})  indicates that
$   V(g)\otimes [V^{-1}(g)]^T M_\omega = A(g) M_\omega,$
where $A(g) = \alpha(g)e^{-\i\theta(g)}$ is a 1-dimensional (1D) linear representation (rep) of $G$. Above equation indicates that $M_\omega$ is the Clebsch-Gordan coefficient that couples the direct product rep $V(g)\otimes [V^{-1}(g)]^T$ to the 1D rep $A(g)$.

Firstly, if the product rep $V(g)\otimes [V^{-1}(g)]^T$ does not contain the 1D rep $A(g)$ at all, then $M_\omega$ must be zero, indicating that the MPS and the reference state are orthogonal. For instance, if the MPS is a product state and if $A(g)\neq 1$ for some $g\in G$, then $M_\omega=0$. In this case, all correlation functions (including the strange correlator) are vanishing. 

Then we consider the case in which the 1D rep $A(g)$ is indeed contained in the product rep $V(g)\otimes [V^{-1}(g)]^T$. Since the character of the direct product rep is real for finite groups or compact Lie groups, ${\rm Tr\ }\big[V(g)\otimes [V^{-1}(g)]^T\big] = |{\rm Tr\ }V(g)|^2$, so if $A(g)$ is contained in the direct product rep, then $A(g)$ must also be real, namely, $A(g)=\pm 1$. Therefore, one must have $A^N(g)=1$ for all $g\in G$ with $N=1$ or $N=2$. 
For finite groups, there must exist an integer $1\leq N\leq |G|$ such that $A^N(g)=1$ for all $g\in G$.  For usual continuous groups having 1D linear reps, $M_\omega\neq0$ generally requires that $A(g)=\pm 1$. Therefore, (\ref{VMV}) indicates that 
$V(g) (M_\omega)^NV^{-1}(g) = (M_\omega)^N, $
for certain $N$, namely, $(M_\omega)^N$ is invariant under the action of the group $G$.

 Hence the degeneracy of the eigenvalues of $(M_\omega)^N$, and then the degeneracy of the absolute value of the eigenvalues of $M_\omega$, are dictated by the dimensionality of the irreducible (projective) reps contained in $V(g)$. Now we analyze in which case $M_\omega$ is norm-degenerate. 

If the MPS is a nontrivial SPT with nontrivial $\beta(g_1,g_2)\neq1,~g_1,g_2\in G$ (here $\beta$ denotes the factor system $\omega$ in Ref.~\cite{chen}, we change the notation here since $\omega$ has been used to denote the trivial product state), then $V(g)$ is a direct sum of irreducible projective reps of the nontrivial class $\beta$, 
$V(g) = \bigoplus_{\nu}  D^{(\nu)}(g),$
with each of the irreducible rep $D^{(\nu)}(g)$ having dimensionality $n_\nu\geq 2$. According to Schur's lemma, the eigenvalue of $(M_\omega)^N$ in each irreducible subspace must be degenerate, just like the entanglement spectrum. Namely, $(M_\omega)^N$ can be written in its eigenspace as
$(M_\omega)^N = \bigoplus_{\nu} \lambda_\nu I_{n_\nu},$
where $I_{n_\nu}$ is the identity matrix of dimension $n_\nu\geq 2$. The degeneracy of the eigenvalues of $(M_\omega)^N$ as well as the entanglement spectrum, is protected by $\beta$ -- the nontrivial class of projective rep. This mechanism confirms the claim in Ref.~\cite{chen} that a nontrivial SPT phase will necessarily lead to a long-range strange correlator. 

Similar arguments holds for SPT protected by time reversal symmetry $T$ or parity $P$. Such nontrivial SPT requires $\beta(T)\ne 1$ or $\beta(P)\ne1$, respectively, following the convention in Ref.~\cite{chen}.
\end{mechanism}

\section{Clebsch-Gordan coefficients and MPS of High Spin AKLT Models}
\label{mps}

In this section, we will provide the expression of the MPS for the Affleck-Kennedy-Lieb-Tasaki (AKLT) models with any integer spin, utilizing the Clebsch-Gordan coefficients.

The construction of AKLT is as follows \cite{AKLT1,book,Hspin2}: We regard each spin-$ S $ site as the direct product of two $ S/2 $ auxiliary spins projected onto the subspace with spin $ S $. The auxiliary spins of adjacent sites form a spin singlet, as shown in Fig.~\ref{fig:1}:

\begin{figure}[h]
	\centering
	\includegraphics[width=0.8\textwidth]{1}
	\caption{AKLT models construction: Two spin-$ S/2 $ particles project onto spin-$ S $ at each site; adjacent sites form spin singlets.}
	\label{fig:1}
\end{figure}

\begin{equation}\begin{aligned}
		\ket{AKLT}:= \otimes^{N+1}P\ket{\alpha}\otimes\ket{bond}^{N}\otimes\ket{\beta}.
\end{aligned}\end{equation}

\noindent Where $ P $ represents the projection onto spin-$ S $ subspace, `bond' refers to the spin singlet state, and $ \ket{\alpha} $ and $ \ket{\beta} $ represent the boundary degrees of freedom.

AKLT state is very suitable to written in MPS form:

\begin{equation}
	\ket{AKLT} = \sum_{\sigma_1,\cdots,\sigma_N}\Tr\left[A^{\left[1\right] }_{\sigma_1}\cdots A^{\left[N\right] }_{\sigma_N} \right] \ket{\sigma_1,\cdots,\sigma_N}.
\end{equation}

From the construction method of AKLT models, it is evident that the tensor product state is a contraction of the projection tensor and the spin singlet: 

\begin{equation}\begin{aligned}
		A^\sigma &= P^\sigma V\\
		A^{\sigma}_{\alpha\beta} &= \sum_\gamma  P^{\sigma}_{\alpha\gamma}V_{\gamma\beta}.
\end{aligned}\end{equation}

$P^{l}_{\alpha\gamma}  $ is the projection tensor $ P^{l}_{\alpha\gamma} := \bra{\alpha,\frac{S}{2};\gamma,\frac{S}{2}}\ket{l,S}$, and $V_{\gamma\beta}$ describes the spin singlet $V_{\gamma\beta} := \bra{\gamma,\frac{S}{2};\beta,\frac{S}{2}}\ket{0,0}$. Both the projection operator and the spin singlet are essentially decompositions of the direct sum of tensor products of representations:

\begin{equation}\begin{aligned}
		\ket{s_3,m_3} = \sum_{m_1=-s_1}^{s_1}  \bra{s_1,m_1;s_2,m_2}\ket{s_3,m_3} \ket{s_1,m_1;s_2,m_2}.
\end{aligned}\end{equation}

\noindent where $ m_i $ ranges from $ -s_i $ to $ s_i $. These coefficients are precisely the Clebsch-Gordan coefficients, whose explicit expressions can be obtained using the Racah formula \cite{Racah}. Consequently, we get Eq.~\eqref{eq:long}.

\begin{equation}\begin{aligned}\label{eq:long}
	&\quad\bra{s_1,m_1;s_2,m_2}\ket{s_3,m_3} =  \delta_{m_3,m_1+m_2}\left[(2s_3+1)\frac{(s_1+s_2-s_3)!(s_2+s_3-s_1)!(s_1+s_3-s_2)!}{(s_1+s_2+s_3+1)!}\right.\\
            &\left.\prod_{i=1}^3(s_i+m_i)!(s_i-m_i)! \right]^{1/2}\times \sum_\nu \left[(-1)^\nu \nu!(s_1+s_2-s_3-\nu)!\right.\\&\left.(s_1-m_1-\nu)!(s_2+m_2-\nu)!(s_3-s_1-m_2+\nu)!(s_3-s_2+m_1+\nu)! \right]^{-1} .
	\end{aligned}\end{equation}

\noindent where the summation for $ \nu $ is constrained to a range that ensures all the factors in the factorials are non-negative. In our case:

\begin{equation}\begin{aligned}
		\widetilde{P}^\sigma_{\alpha\beta} &= \bra{j,\alpha;j,\beta}\ket{2j,\sigma}\\
        &= \delta_{\sigma,\alpha+\beta}\left[ \frac{ (2j)!(2j)!(2j+\sigma)!(2j-\sigma)!  }{(4j)!(j+\alpha)!(j-\alpha)!(j+\beta)!(j-\beta)! }\right]^{1/2}.	
\end{aligned}\end{equation}

\begin{equation}\begin{aligned}
		\widetilde{V}_{\alpha\beta} &= \delta_{\alpha,-\beta}\frac{(-1)^{j-\alpha}}{\sqrt{2j+1}} 
\end{aligned}\end{equation}

It is interesting to mention that $ j,\alpha $ might not be an integer but a half-integer; however, $ j-\alpha $ must be an integer.


After a normalization, the final result is Eq.~\eqref{eq:long2}.

	\begin{equation}\begin{aligned}\label{eq:long2}
			A^\sigma_{\alpha\beta} &= \delta_{\sigma,\alpha-\beta} (-1)^{j+\beta} \left[ \frac{ (2j)!(2j+1)!(2j+\sigma)!(2j-\sigma )!  }{(4j+1)!(j+\alpha)!(j-\alpha)!(j+\beta)!(j-\beta)! }\right]^{1/2}	.
	\end{aligned}\end{equation}

In addition, it can be proved that $A^\sigma_{\alpha\beta}$ is a bicanonical form.

\section{Proof of Theorems and Claims}\label{pot}
\subsection{Theorem 1}\label{stricter}
In the main text, we focus on the \textit{connected strange correlator}. Analytic analysis (\textit{Theorem} \ref{magnitude-deg}, \textit{Mechanism} \ref{SPT}) has shown that it is the connected strange correlator that reflects the contribution from the nontrivial SRE state. Since we are discussing spurious signature of the original strange correlator, we show in this section that the connected strange correlator provides a stricter test of long-range order than the strange correlator. That is to say, if we can find a long-range connected strange correlator for a MPS target state, then so can we find a long-range strange correlator, and the reverse is not true.

The proof of \textit{Theorem \ref{theorem3}} is based on the proof of \textit{Theorem \ref{magnitude-deg},\ref{theorem2}}, but appears first in the main text. Readers may read SM Sec.~\ref{tm1} and \ref{tm2} first for preliminaries.

\begin{theorem}

Given a Matrix Product State (MPS) target state $|\Phi\rangle$ and a reference state $|\Omega\rangle$. If there exist local operators $O^a$ and $O^b$ such that the connected strange correlator exhibits long-range behavior in the thermodynamic limit, i.e.,
\begin{equation}
    \lim_{|i-j|\to\infty} \langle O^a_i O^b_j \rangle_{S, \text{connected}} \neq 0,
\end{equation}
then there exist (in fact, almost all) operators $\tilde{O}^a$ and $\tilde{O}^b$  such that the ordinary strange correlator also exhibits long-range behavior:
\begin{equation}
    \lim_{|i-j|\to\infty} \langle \tilde{O}^a_i \tilde{O}^b_j \rangle_{S} \neq 0.
\end{equation}
\end{theorem}
\begin{proof}
The proof follows directly from the spectral decomposition of the transfer matrix $M_\omega$ established in \textit{Theorem \ref{magnitude-deg}} and the genericity argument in \textit{Theorem \ref{theorem2}}.

By the contrapositive of the first part of \textit{Theorem \ref{magnitude-deg}} (see Eq.~\eqref{csc} and associated proof in App.~\ref{tm1}), if the largest-norm eigenvalue of the transfer matrix $M_\omega$ is non-degenerate (i.e., degeneracy $D=1$), then the connected strange correlator vanishes for \textit{all} possible operators $O^a, O^b$. 

Therefore, the premise that there exists at least one pair of operators $O^a, O^b$ yielding a non-zero long-range connected strange correlator implies that the largest-norm eigenvalue of $M_\omega$ must be magnitude-degenerate with $D > 1$. That is:
\begin{equation}
    |\lambda_0| = |\lambda_1| = \dots = |\lambda_{D-1}| > |\lambda_D|, \quad \text{with } D \ge 2.
\end{equation}

Having established that $D > 1$, we examine the behavior of the ordinary strange correlator. In the thermodynamic limit ($L\to\infty$) followed by the large distance limit ($R \equiv |i-j| \to \infty$), the ordinary strange correlator takes the asymptotic form (derived in Eq.~\eqref{sc_2}):
\begin{equation}
    \langle \tilde{O}^a \tilde{O}^b \rangle_{S} \sim \frac{\lambda_0^{-2} \sum_{i,j < D} \tilde{C}_{ij} e^{i[\theta_i R + \theta_j(L-R-2)]}}{\sum_{k < D} e^{i \theta_k L}},
\end{equation}
where $\tilde{C}_{ij} = \text{Tr}[\tilde{M}_a P_i \tilde{M}_b P_j]$ depends on the choice of operators $\tilde{O}^a, \tilde{O}^b$.

For the ordinary strange correlator to vanish (or decay exponentially) for \textit{all} operators, the coefficients $\tilde{C}_{ij}$ would need to satisfy specific cancellation conditions for every possible choice of $\tilde{O}^a$ and $\tilde{O}^b$. However, as proven in \textit{Theorem \ref{theorem2}} (App.~\ref{tm2}), the set of operator matrices $(\tilde{M}_a, \tilde{M}_b)$ for which the strange correlator vanishes forms a set of zero Lebesgue measure in the operator space $\mathcal{M}$ when $D > 1$.

Since the operator space $\mathcal{M}$ is not empty, there necessarily exist operators $\tilde{O}^a, \tilde{O}^b$ (in fact, almost all local operators) that lie outside this zero-measure set. For these operators, the sum in the numerator does not vanish, resulting in long-range behavior (either saturation to a constant or persistent oscillation). 

Thus, the existence of a long-range connected strange correlator guarantees the existence of operators yielding a long-range ordinary strange correlator. In fact, the theorem can be strengthened to ``the existence of a long-range connected strange correlator guarantees almost all operators yielding a long-range ordinary strange correlator'',  as proved above. 

One can immediately verify that the reverse is not true. As shown in {Eq.~\eqref{tps}} in the main text, two trivial product states can yield nonzero strange correlators due to the symmetry breaking nature. However, since the MPS representation of product states always have $D=1$ bond dimension, therefore the magnitude-degeneracy cannot happen since there are only 1 eigenvalue. According to \textit{Theorem~\ref{magnitude-deg}}, this always gives a zero connected strange correlator, regardless of the operator choice.
\end{proof}

\subsection{Theorem 2}\label{tm1}
\begin{theorem}[Magnitude-degeneracy implies long-range strange correlator]\label{magnitude-deg}
Suppose the eigenvalue of $M_\omega$ with the largest magnitude is $D$-fold magnitude-degenerate. 
i.e. we assume that
   $ |\lambda_0|=|\lambda_1|=\cdots=|\lambda_{D-1}|>|\lambda_{D}|>\cdots.$

If $D=1$, then $\forall O^a,O^b$, the connected strange correlator is zero:
\begin{equation}\label{csc}
    \expval{O^aO^b}_{S,\text{connected}}:=\expval{O^aO^b}_S-\expval{O^a}_S\expval{O^b}_S=0.
\end{equation}

    If $D>1$, then there exist some operator $O^a,O^b$  and an infinite sequence $\{R_i\}:=R_1,R_2,\cdots$ where $j>i\iff R_j>R_i>0$ and $\lim_{i\to\infty} R_i=\infty$, such that
\begin{equation}
    \lim_{k\to \infty} \expval{O^a_iO^b_{i+R_k+1}}_{S,\text{connected}}\ne 0.
\end{equation}
\end{theorem}

\textit{Theorem \ref{magnitude-deg}} highlights the relation between the magnitude degeneracy of transfer matrix and the presence of the long-range connected strange correlator. The proof is presented below.

\begin{proof}
In the calculation of strange correlators, we first take the $L\to\infty$ limit, and take $R\to\infty$ afterwards. For simplicity we abbreviate $\lim_{R\to\infty}\lim_{L\to\infty}$ as $\lim_{R,L\to \infty}$ from now on.

The matrix $ M_\omega $ is decomposed into $M_\omega = \sum_{i=0} \lambda_i P_i$, with $\lambda_i$ representing the eigenvalue and $P_i$ serving as the associated projection operator, whereby we assume $ \abs{\lambda_0}\ge\abs{\lambda_1}\ge\abs{\lambda_2}\cdots$. $P_iP_j=\delta_{ij}P_j$ holds naturally for projectors.

According to Eq.~\eqref{sc_def}, in the thermodynamic limit, we have
\begin{equation}
    \begin{aligned}
        &\lim_{R,L\to\infty}\expval{O^a_iO^b_{i+R+1}}_S\\ &= \lim_{R,L\to\infty}\frac{\Tr\left[M_a M_\omega^R M_b M_\omega^{L-R-2} \right]}{\Tr\left[M_\omega^L \right]}\\
        &=\lim_{R,L\to\infty}\frac{\Tr\left[M_a (\sum_{i}\lambda_iP_i)^{R}  M_b (\sum_{j}\lambda_jP_j)^{L-R-2} \right]}{\Tr\left[(\sum_{k}\lambda_kP_k)^L \right]}
\end{aligned}
\end{equation}

$P_i$'s are projectors, which means
\begin{equation}
    P_iP_j=\delta_{ij}P_i.
\end{equation}
Using this fact, the limit above becomes

\begin{equation}
\begin{aligned}
        &\lim_{R,L\to\infty}\frac{\lambda_0^{-L}\Tr\left[M_a (\sum_{i}\lambda_iP_i)^{R} M_b (\sum_{j}\lambda_jP_j)^{L-R-2} \right]}{\lambda_0^{-L}\Tr\left[\sum_{k}\lambda_k^LP_k \right]}\\
        &=\lim_{R,L\to\infty}\frac{\lambda_0^{-L}\Tr\left[M_a (\sum_{i}\lambda_i^{R}P_i) M_b \left(\sum_{j}\lambda_j^{L-R-2}P_j\right) \right]}{\lambda_0^{-L}\left(\sum_{k}\lambda_k^L\right)\Tr[P_0]}\\
        &=\lim_{R,L\to\infty}\frac{\lambda_0^{-L}\sum_i \lambda_i^R\sum_{j} \lambda_j^{L-R-2}\Tr\left[M_a P_i M_b  P_j \right]}{\sum_{i}\left(\frac{\lambda_i}{\lambda_0}\right)^L\Tr[P_0]}.
    \end{aligned}
\end{equation}

The denominator depends on the ratio of $\lambda_i$ and $\lambda_0$. 
Here naturally  $|\lambda_0|>0$ otherwise the matrix is null. We define that
\begin{equation}
   a_ie^{i\theta_i}:= \frac{\lambda_i}{\lambda_0}
\end{equation}
and
\begin{equation}
\begin{aligned}
C_{i}^a&:=\Tr[P_iM_a]/\Tr[P_0]\\
    C_{ij}^{ab}&:=\Tr[M_aP_iM_bP_j]/\Tr[P_0]
\end{aligned}
\end{equation}
then the limit above eventually simplifies to
\begin{equation}
    \begin{aligned}
&\lim_{R,L\to\infty}\expval{O^a_iO^b_{i+R+1}}_S\\&=\lim_{R,L\to\infty}\frac{\lambda_0^{-2}\sum_i C_{ij}^{ab}\left(\frac{\lambda_i}{\lambda_0}\right)^R\left(\frac{\lambda_j}{\lambda_0}\right)^{L-R-2}}{\sum_{i}\left(\frac{\lambda_i}{\lambda_0}\right)^L}.
    \end{aligned}
\end{equation}
Similarly we obtain the expression:
\begin{equation}
   \lim_{L\to\infty} \expval{O^a_i} = \lim_{L\to\infty}\frac{\lambda_0^{-1}\sum_i C_{i}^{a}\left(\frac{\lambda_i}{\lambda_0}\right) ^{L-1}}{\sum_{i}\left(\frac{\lambda_i}{\lambda_0}\right)^L}.
\end{equation}
After the limit is taken, suppose the largest eigenvalue of $M_\omega$ is $D$-fold magnitude degenerate, we have
\begin{equation}
    \lim_{n\to\infty}\left(\frac{\lambda_i}{\lambda_0}\right)^n=\begin{cases}
        0,&i\ge D;\\
        \lim_{n\to\infty}e^{\i n\theta_i},&0\le i<D.
    \end{cases}
\end{equation}
and $\theta_0=0$ by definition.

This property indicates that the thermodynamic limit acts effectively as truncating all eigenvalues whose magnitude is smaller than the largest-magnitude eigenvalue. Therefore the summation in the limit above can be simplified. It follows that
\begin{equation}\label{sc_2}
    \begin{aligned}
        &\lim_{R,L\to\infty}\expval{O^a_iO^b_{i+R+1}}_S\\&=\lim_{R,L\to\infty}\frac{\lambda_0^{-2} \sum_{i,j<D}C_{ij}^{ab}e^{\i [\theta_i R+\theta_j(L-R-2)]}}{\sum_{i<D}e^{\i\theta_iL}}
    \end{aligned}
\end{equation}
and
\begin{equation}\label{sev_2}
    \lim_{L\to\infty} \expval{O^a_i} = \lim_{L\to\infty}\frac{\lambda_0^{-1}\sum_{i<D} C_{i}^{a}e^{\i\theta_i(L-1)}}{\sum_{i<D} e^{\i\theta_iL}}.
\end{equation}
Now we can discuss the correspondence between the magnitude degeneracy and the long-range behavior of strange correlator.

\textit{(i) If the largest eigenvalue is non-degenerate,} i.e. $D=1$, then this yields
\begin{equation}
    \lim_{R,L\to\infty}\expval{O^a_iO^b_{i+R+1}}_S= \lambda_0^{-2}C_{00}^{ab}
\end{equation}
and
\begin{equation}
    \lim_{L\to\infty} \expval{O^a_i}_S = \lambda_0^{-1}C_0^a.
\end{equation}
This yields
\begin{equation}\label{sc_1}
    \lim_{R,L\to\infty}\expval{O^aO^b}_{S,\text{connected}}=\lambda_0^{-2}(C_{00}^{ab}-C_0^aC_0^b).
\end{equation}
Having assumed the non-degeneracy, the rank of projector $P_0$ is 1, this gives
\begin{equation}\label{simplify}
    \Tr[M_aP_0M_bP_0]=\Tr[P_0M_aP_0]\Tr[P_0M_bP_0]
\end{equation}
and thus
\begin{equation}\label{non-degen}
    \expval{O^aO^b}_{S,\text{connected}}=0.
\end{equation}
This proves that the absence of degeneracy in the transfer matrix implies the absence of long-range strange correlator. 

In the case of AKLT models, the operators $O^aO^b$ are usually taken as $S^+_iS^-_j$ and $\ket{\Omega}=\ket{0\cdots0}$, which yields $\expval{S^\pm}_S=0$, thus we have
\begin{equation}
    \expval{S^+_iS^-_j}_{S,\text{connected}}=\expval{S^+_iS^-_j}_{S}
\end{equation}
in this case.

\textit{(ii) If the largest eigenvalue is magnitude-degenerate,} i.e. $D>1$, then the case becomes more complicated. While the long-range correlation is expected here, the phase factor in Eq.~\eqref{sc_2} and Eq.~\eqref{sev_2} may not have a well-defined limit, so we turn to prove that the correlation function does not decay to zero. This includes the normally expected constant nonzero strange correlators and the case that the correlation function fluctuates and does not converge when $R\to\infty$. In both cases, we can say that the system exhibits long-range behavior since we can always pick an arbitrarily large $R$ such that the strange correlator is still nonzero.

The proof is completed by contradiction. First we assume otherwise, i.e. we assume that $\forall O_a,O_b,\{R_i\}$, $\lim_{k\to \infty} \expval{O^a_iO^b_{i+R_k+1}}_{S,\text{connected}}=0$.
Notice that the expression of strange correlators explicitly depends on $R$ and $L$.

\begin{lemma}[Ces\`aro Mean \cite{rudin1976principles}]\label{cesaro}
    If  an infinite sequence $\{z_i\},~z_i\in\mathbb C$ satisfies
    \begin{equation}
        \lim_{i\to\infty} z_i=Z,
    \end{equation}
    then 
    \begin{equation}
        \lim_{i\to\infty} \frac{1}{i}\sum_{n=1}^iz_n=Z.
    \end{equation}
\end{lemma}

Here, using the fact that we have assumed the limit of the connected strange correlator is zero, we first simplify that
\begin{equation}
    \begin{aligned}
         &\lim_{R,L\to\infty}\expval{O^aO^b}_S-\expval{O^a}\expval{O^b}\\
  &=\lim_{R,L\to\infty}{\lambda_0^{-2} }\left[\frac{\sum_{i,j<D}C_{ij}^{ab}e^{\i [(\theta_i-\theta_j )R+\theta_j(L-2)]}}{\sum_{i<D}e^{\i\theta_iL}}-\frac{\sum_{i,j<D} C_{i}^{a}C_j^be^{\i(\theta_i+\theta_j)(L-1)}}{(\sum_{i<D} e^{\i\theta_iL})^2}\right]\\
 &=\lim_{R,L\to\infty}\frac{(\sum_{i<D} e^{\i\theta_iL})^{-2}}{\lambda_0^{2} }\left[{\sum_{i,j,k<D}C_{ij}^{ab}e^{\i [(\theta_i-\theta_j )R+(\theta_j+\theta_k)L]}}e^{-2\i\theta_j}-{\sum_{i,j<D} C_{i}^{a}C_j^be^{\i(\theta_i+\theta_j)(L-1)}}\right].
    \end{aligned}
\end{equation}
We notice that the coefficient
$(\sum_{i<D} e^{\i\theta_iL})^{-2}{\lambda_0^{-2} }$
does not converge to zero for any configurations of $\theta_i$ since
\begin{equation}
    \left|\frac{(\sum_{i<D} e^{\i\theta_iL})^{-2}}{\lambda_0^{2} }\right|\ge (D|\lambda_0|)^{-2}>0,
\end{equation}
and this lower bound is independent of $L$. Therefore we conclude that
\begin{equation}
    \begin{aligned}
        &\lim_{R,L\to\infty}\left[{\sum_{i,j,k<D}C_{ij}^{ab}e^{\i [(\theta_i-\theta_j )R+(\theta_j+\theta_k)L]}}e^{-2\i\theta_j}-{\sum_{i,j<D} C_{i}^{a}C_j^be^{\i(\theta_i+\theta_j)(L-1)}}\right]=0.
    \end{aligned}
\end{equation}
It follows that the following limit is also zero:
\begin{align}
    \begin{aligned}
        &\lim_{R,L\to\infty}\frac{1}{L}\sum_{l=1}^L\left[{\sum_{i,j,k<D}C_{ij}^{ab}e^{\i [(\theta_i-\theta_j )R+(\theta_j+\theta_k)l]}}e^{-2\i\theta_j}-{\sum_{i,j<D} C_{i}^{a}C_j^be^{\i(\theta_i+\theta_j)(l-1)}}\right]\\
 =&\lim_{R\to\infty}\left[{\sum_{i,j,k<D}C_{ij}^{ab}e^{\i [(\theta_i-\theta_j )R]}}e^{-2\i\theta_j}\delta_{\theta_j+\theta_k,0}-{\sum_{i,j<D} C_{i}^{a}C_j^b\delta_{\theta_{i}+\theta_j,0}}\right]\\
\end{aligned}
\end{align}
Now we apply \textit{ Lemma \ref{cesaro}}. Here the role of Ces\`aro mean effectively eliminates all oscillating terms, which yields
\begin{align}
    \begin{aligned}
  =&\lim_{R\to\infty}\frac{1}{R}\sum_{r=1}^R\left[{\sum_{i,j,k<D}C_{ij}^{ab}e^{\i [(\theta_i-\theta_j )r]}}e^{-2\i\theta_j}\delta_{\theta_j+\theta_k,0}-{\sum_{i,j<D} C_{i}^{a}C_j^b\delta_{\theta_{i}+\theta_j,0}}\right]\\
 =&{\sum_{i,j,k<D}C_{ij}^{ab}}e^{2\i\theta_k}\delta_{\theta_j+\theta_k,0}\delta_{\theta_i,\theta_j}-{\sum_{i,j<D} C_{i}^{a}C_j^b\delta_{\theta_{i}+\theta_j,0}}\\
 = &~0.
\end{aligned}
\end{align}

This is equivalent to assuming that
\begin{equation}\label{degen-cond}
\begin{aligned}
    \sum_{i,j,k<D} \left(\frac{\lambda_k}{\lambda_0}\right)^2\Tr[M_aP_iM_bP_j]\delta_{\theta_j+\theta_k,0}\delta_{\theta_i,\theta_j}\\=\sum_{i,j<D}\Tr[M_aP_i]\Tr[M_bP_j]\delta_{\theta_i+\theta_j,0}
\end{aligned}
\end{equation}
$\forall M_a, M_b$. When $D=1$, this condition naturally reduces to Eq.~\eqref{simplify} and is automatically satisfied. However, having assumed that $D>1$, since $O^a,O^b$ can be arbitrarily chosen, and the equation above does not automatically hold for every $M_a, M_b$ for $D>1$, so we can always chose some $M_a, M_b$ such that the original connected strange correlator does not converge to zero, which is contradictory to our assumption. Actually, this condition equivalently ascertained a hyperplane in the space of $M_a\oplus M_b$, so we can always choose $O^a, O^b$ such that the strange correlator exhibit long-range behavior.

Here we have finished our proof for \textit{Theorem \ref{magnitude-deg}}.
\end{proof}

\subsection{Theorem 3}\label{tm2}

\begin{theorem}
    If $D>1$, we define the direct sum of spaces of operator matrices $M_a, M_b$ as $\mathcal M:=\{(M_a,M_b)|M_a := \sum_\sigma \bra{\omega}O^a\ket{\sigma} M^{\sigma} ,
		M_b := \sum_\sigma \bra{\omega}O^b\ket{\sigma} M^{\sigma}\}$, then $\expval{O^a_iO^b_j}_{S,\text{connected}}$ does not converge to zero (i.e. exhibits long-range behavior) almost everywhere in $\mathcal M$, where the exceptions have zero Lebesgue measure in $\mathcal M$.
\end{theorem}

\textit{Theorem \ref{theorem2}} claims that if the largest-magnitude eigenvalue is magnitude-degenerate, then almost all local operators can yield long-range connected strange correlator in the thermodynamic limit. The proof is presented below.

\begin{proof}
    First we prove that Eq.~\eqref{degen-cond} is not an identity when $D>1$. This can be done by providing a counterexample. We let $D=2$ and $\lambda_0=\lambda_1$ and $M_a=P_0,~M_b=P_1$. Here Eq.~\eqref{degen-cond} becomes
    \begin{equation}
       2C_{00}^{ab}+2C_{11}^{ab}+2C_{10}^{ab}+2C_{01}^{ab}=C_0^aC_1^b+C_1^aC_0^b+C_0^aC_0^b+C_1^aC_1^b.
    \end{equation}
    Using $P_iP_j=\delta_{ij}$, this becomes
    \begin{equation}
        0=\Tr[P_0]\Tr[P_1]
    \end{equation}
    which simplifies to 
    \begin{equation}
        0=1,
    \end{equation}
    leading to a contradiction. Therefore we conclude that the Eq.~\eqref{degen-cond} is not automatically satisfied.

    Now we define the space $(M_a,M_b)\in  \cal M$, which is the space of the direct sum of $M_a$ and $M_b$. If $M_a$ and $M_b$ are $N\times N$ matrices with complex entries, then $\mathcal{M} \cong \mathbb R^{4N^2}$. 
    We define a function $F:\mathcal M\to \mathbb C$, which writes
    \begin{equation}
        \begin{aligned}
    F(M_a,M_b)&=\sum_{i,j,k<D} \left(\frac{\lambda_k}{\lambda_0}\right)^2\Tr[M_aP_iM_bP_j]\delta_{\theta_j+\theta_k,0}\delta_{\theta_i,\theta_j}\\&-\sum_{i,j<D}\Tr[M_aP_i]\Tr[M_bP_j]\delta_{\theta_i+\theta_j,0}.
\end{aligned}
    \end{equation}
Now we let $F_1=\Re{F}$ be the real part of $F$, which is a map $F_1:\mathbb{R}^{4N^2}\to \mathbb R$. Let $S=\{(M_a,M_b)|F(M_a,M_b)=0\}$ be the zero point set of $F$ and $S_1=\{(M_a,M_b)|F_1(M_a,M_b)=0\}$ be the zero point set of $F_1$, then it is obvious that $S\subseteq S_1$. $S$ corresponds to the set of $(M_a,M_b)$ that satisfies the constraint given by Eq.~\eqref{degen-cond}. Since $F$ is essentially some polynomials of entries of $M_a$ and $M_b$, $F$ is analytic, so $F_1$ is also analytic. We have already shown that $F$ is not identically zero, and without loss of generality we let $F_1$ is not identically zero. 

We have the following useful lemma in real analysis:
\begin{lemma}[Zero sets of real analytic functions \cite{mityagin2015zerosetrealanalytic}]\label{zero-point}
    Let $F$ be a real analytic function on (a connected open domain $U$ of) $\mathbb R^d$. If $F$ is not identically zero, the its zero set
    \begin{equation}
        S(F):=\{x\in U|F(x)=0\}
    \end{equation}
    has a zero Lebesgue measure, i.e.
    \begin{equation}
        \operatorname{mes}_dS(F)=0,
    \end{equation}
    where $\operatorname{mes}_d$ denotes the Lebesgue measure in $\mathbb R^d$.
    
\end{lemma}
    This immediately gives that
    \begin{equation}
        \operatorname{mes}_{4N^2}S_1=0,
    \end{equation}
    and since $S\subseteq S_1$,
    \begin{equation}
        \operatorname{mes}_{4N^2}S=0.
    \end{equation}
    \end{proof}

\subsection{Symmetry Breaking Case}\label{ssb-proof}
In the main text, we claim that if the MPS target state is the symmetric ground state of a symmetry breaking Hamiltonian (this state can be selected by linear combining bases of all ground state subspaces), then we can always choose the reference state $\ket{\Omega}=\bigotimes_i \ket\omega_i$, such that the largest-magitude eigenvalue of the transfer matrix $M_\omega=M_i\omega_i^*$ is magnitude-degenerate.

Suppose the whole symmetry group of the system Hamiltonian is $G$, and a subgroup $K \subset G$ is spontaneously broken. It follows that the symmetry action $u(k)$ for a specific $k \in K$ maps a ground state $|\psi_0\rangle$ to another orthogonal ground state $|\psi_1\rangle$. Without loss of generality:
\begin{eqnarray}\label{sym-effect}u(k)|\psi_0\rangle = e^{i\phi_k}|\psi_1\rangle.\end{eqnarray}
Following Ref.~\cite{chen}, the symmetric ground state will have a block-diagonalized MPS form:
\begin{eqnarray}
M_i = \begin{pmatrix}M^{(0)}_i & & \\& M^{(1)}_i & \\&& \ddots
\end{pmatrix}
\end{eqnarray}
with the number of blocks greater than 1, and the double tensor $\mathbb{E}^{(n)} = \sum_{i} M_i^{(n)} \otimes M_i^{(n)*}$ for each block sharing the same largest eigenvalue. This forces the mixed transfer matrix to share the same block-diagonal form:\begin{equation}M_\omega = \sum_i M_i \omega_i^* = \begin{pmatrix}M^{(0)}_\omega &  & \\& M^{(1)}_\omega & \\&& \ddots\end{pmatrix},\end{equation}where $M^{(n)}_\omega = \sum_i M^{(n)}_i \omega_i^*$. Given that $|\psi_n\rangle = \sum_{\{i\}} \text{Tr}[M_{i_1}^{(n)}\cdots M_{i_N}^{(n)}] |i_1\cdots i_N\rangle$, Eq.~\eqref{sym-effect} implies the following relation between the MPS tensors of different blocks (see Ref.~\cite{chen} for details):
\begin{eqnarray}\label{mps-block-relation}\sum_{j} u(k)_{ij} M_j^{(0)} = e^{i\phi_k} V^{-1}(k) M_i^{(1)} V(k),\end{eqnarray}
where $V(k)$ is a gauge transformation matrix. To relate the transfer matrices $M^{(0)}_\omega$ and $M^{(1)}_\omega$, we assume the reference state $|\Omega\rangle = \bigotimes_{site} (\sum_i \omega_i |i\rangle)$ preserves the symmetry $u(k)$, satisfying $u(k)|\Omega\rangle = e^{i\theta_k}|\Omega\rangle$. At the single-site level, this implies:\begin{equation}\label{ref-sym}\sum_{j} u(k)_{ij} \omega_j = e^{i\theta_k} \omega_i \implies \sum_{i} \omega_i^* u(k)_{ji}^\dagger = e^{-i\theta_k} \omega_j^*.\end{equation}
Starting from the definition of $M^{(0)}_\omega$ and inserting the identity $\sum_p u(k)_{jp}^\dagger u(k)_{pi} = \delta_{ji}$:\begin{equation}\begin{aligned}M^{(0)}_\omega&= \sum_j M_j^{(0)} \omega_j^* \\&= \sum_j \left( \sum_p u(k)_{jp}^\dagger \sum_i u(k)_{pi} M_i^{(0)} \right) \omega_j^* \\ &= \sum_p \left( \sum_j \omega_j^* u(k)_{jp}^\dagger \right) \left( \sum_i u(k)_{pi} M_i^{(0)} \right).\end{aligned}\end{equation}
Substituting the MPS transformation rule Eq.~\eqref{mps-block-relation} and the reference state symmetry Eq.~\eqref{ref-sym} into the equation above yields:
\begin{equation}\begin{aligned}M^{(0)}_\omega &= \sum_p \left( e^{-i\theta_k} \omega_p^* \right) \left( e^{i\phi_k} V^{-1}(k) M_p^{(1)} V(k) \right)\\ &= e^{i(\phi_k - \theta_k)} V^{-1}(k) \left( \sum_p M_p^{(1)} \omega_p^* \right) V(k) \\ &= e^{i(\phi_k - \theta_k)} V^{-1}(k) M^{(1)}_\omega V(k).\end{aligned}\end{equation}
Thus, $M^{(0)}_\omega$ and $M^{(1)}_\omega$ differ only by a similarity transformation and a global phase shift. Consequently, they share the exact same eigen-spectrum magnitude, leading to the magnitude-degeneracy of the total transfer matrix $M_\omega$.

\section{More Examples of Spurious Strange Correlators}\label{more-examples}

\setcounter{example}{3}
 The argument in \textit{Mechanism~\ref{phase-mismatch}} can be also applied to the case of parity symmetry, where the parity transformation $P=w\otimes w\otimes\cdots \otimes w P_1$ with $w$ being the onsite unitary transformation satisfying $w^2=1$ and $P_1$ being the exchange of sites.
\begin{example}[Parity phase mismatch]\label{parity-mismatch}
    As a concrete example, we consider an MPS model where both the physical spin and virtual spin are 1, and its MPS representation can be written as:

\begin{equation}
\begin{aligned}
    	M^{\left[ -1\right] } &= \begin{pmatrix}0&-1&0\\
	0&0&-1\\0&0&0\end{pmatrix},\\M^{\left[ 0\right] } &= \begin{pmatrix}1&0&0\\
	0&0&0\\0&0&-1\end{pmatrix},\\M^{\left[ +1\right] } &= \begin{pmatrix}0&0&0\\
	1&0&0\\0&1&0\end{pmatrix}
\end{aligned}
\end{equation}

We can check that, under spatial inversion with center on the bonds (onsite unitary set to be identity $I$, which means we only change the position of sites), we have $ M^{\sigma T} = - V^{-1}M^\sigma V$, and $V^T = \beta V =V$. So $\alpha(P)=-1,~\beta(P)=1$ (definitions follow Ref.~\cite{chen}, check App.~\ref{review} for a brief review).

Similar to our previous example, we construct the following $SO(3)$-invariant spin one-body state: $(\ket{1,-1} - \ket{0,0} + \ket{-1,1})/\sqrt{3}$ which preserves $SO(3)$ symmetry, and here $e^{\i\theta(g)}=1$. Here again we have merged two adjacent sites into a new site to obtain a symmetric reference state. The strange correlator can then be obtained, with the transfer matrix $M_\omega$ being $-\frac{2}{\sqrt{3}}\mathbb{I}$, explicitly showing degeneracy. Here, a single-site operator corresponds to two previous operators since we have merged two sites into one. We pick $O^a=S^+\otimes I,~O^b=I\otimes S^-$, then
\begin{eqnarray}
    \consc{O^aO^b}=\sev{O^aO^b}=-\frac{1}{3},
\end{eqnarray}
explicitly demonstrating long-range strange correlator.
\end{example}

There are also examples where different mechanisms exist simultaneously.
\begin{example}[Spin-2 AKLT with phase mismatch]\label{mixed-example}
    For $S=2$ AKLT model, the spin rotation around $y$-axis by angle $\pi$, $R_y:=e^{\i\pi S_y}$, act effectively as
\begin{equation}
    R_y=\begin{pmatrix}0&0&0&0&1\\0&0&0&-1&0\\0&0&1&0&0\\0&-1&0&0&0\\1&0&0&0&0\end{pmatrix}.
\end{equation}

For spin-2 AKLT model, $\alpha(e^{\i\pi S_y})=0$. If we pick $\ket{\omega}=\sum_i\omega_i\ket{i}$ as $\frac{1}{\sqrt{2}}(0,1,0,1,0)^T$ which satisfies $R_y\ket{\omega}=-\ket{\omega}$, and $O^a_iO^b_j=S^+_iS^-_j$, then the strange correlator exhibits long-range behavior, even though the system belongs to the trivial SPT phase. Here we can also pick $O^a_iO^b_j=S^+_iS^-_j$, yielding
\begin{eqnarray}
    \consc{S^+S^-}=\sev{S^+S^-}=-4.
\end{eqnarray}
\end{example}


\begin{thebibliography}{74}%
\makeatletter
\providecommand \@ifxundefined [1]{%
 \@ifx{#1\undefined}
}%
\providecommand \@ifnum [1]{%
 \ifnum #1\expandafter \@firstoftwo
 \else \expandafter \@secondoftwo
 \fi
}%
\providecommand \@ifx [1]{%
 \ifx #1\expandafter \@firstoftwo
 \else \expandafter \@secondoftwo
 \fi
}%
\providecommand \natexlab [1]{#1}%
\providecommand \enquote  [1]{``#1''}%
\providecommand \bibnamefont  [1]{#1}%
\providecommand \bibfnamefont [1]{#1}%
\providecommand \citenamefont [1]{#1}%
\providecommand \href@noop [0]{\@secondoftwo}%
\providecommand \href [0]{\begingroup \@sanitize@url \@href}%
\providecommand \@href[1]{\@@startlink{#1}\@@href}%
\providecommand \@@href[1]{\endgroup#1\@@endlink}%
\providecommand \@sanitize@url [0]{\catcode `\\12\catcode `\$12\catcode `\&12\catcode `\#12\catcode `\^12\catcode `\_12\catcode `\%12\relax}%
\providecommand \@@startlink[1]{}%
\providecommand \@@endlink[0]{}%
\providecommand \url  [0]{\begingroup\@sanitize@url \@url }%
\providecommand \@url [1]{\endgroup\@href {#1}{\urlprefix }}%
\providecommand \urlprefix  [0]{URL }%
\providecommand \Eprint [0]{\href }%
\providecommand \doibase [0]{https://doi.org/}%
\providecommand \selectlanguage [0]{\@gobble}%
\providecommand \bibinfo  [0]{\@secondoftwo}%
\providecommand \bibfield  [0]{\@secondoftwo}%
\providecommand \translation [1]{[#1]}%
\providecommand \BibitemOpen [0]{}%
\providecommand \bibitemStop [0]{}%
\providecommand \bibitemNoStop [0]{.\EOS\space}%
\providecommand \EOS [0]{\spacefactor3000\relax}%
\providecommand \BibitemShut  [1]{\csname bibitem#1\endcsname}%
\let\auto@bib@innerbib\@empty
\bibitem [{\citenamefont {Chen}\ \emph {et~al.}(2011{\natexlab{a}})\citenamefont {Chen}, \citenamefont {Gu},\ and\ \citenamefont {Wen}}]{chen}%
  \BibitemOpen
  \bibfield  {author} {\bibinfo {author} {\bibfnamefont {X.}~\bibnamefont {Chen}}, \bibinfo {author} {\bibfnamefont {Z.-C.}\ \bibnamefont {Gu}},\ and\ \bibinfo {author} {\bibfnamefont {X.-G.}\ \bibnamefont {Wen}},\ }\bibfield  {title} {\bibinfo {title} {Complete classification of one-dimensional gapped quantum phases in interacting spin systems},\ }\href {https://doi.org/10.1103/PhysRevB.84.235128} {\bibfield  {journal} {\bibinfo  {journal} {Phys. Rev. B}\ }\textbf {\bibinfo {volume} {84}},\ \bibinfo {pages} {235128} (\bibinfo {year} {2011}{\natexlab{a}})}\BibitemShut {NoStop}%
\bibitem [{\citenamefont {Affleck}\ \emph {et~al.}(1987)\citenamefont {Affleck}, \citenamefont {Kennedy}, \citenamefont {Lieb},\ and\ \citenamefont {Tasaki}}]{AKLT1}%
  \BibitemOpen
  \bibfield  {author} {\bibinfo {author} {\bibfnamefont {I.}~\bibnamefont {Affleck}}, \bibinfo {author} {\bibfnamefont {T.}~\bibnamefont {Kennedy}}, \bibinfo {author} {\bibfnamefont {E.~H.}\ \bibnamefont {Lieb}},\ and\ \bibinfo {author} {\bibfnamefont {H.}~\bibnamefont {Tasaki}},\ }\bibfield  {title} {\bibinfo {title} {{Rigorous results on valence-bond ground states in antiferromagnets}},\ }\href {https://doi.org/10.1103/PhysRevLett.59.799} {\bibfield  {journal} {\bibinfo  {journal} {Phys. Rev. Lett.}\ }\textbf {\bibinfo {volume} {59}},\ \bibinfo {pages} {799} (\bibinfo {year} {1987})}\BibitemShut {NoStop}%
\bibitem [{\citenamefont {Haldane}(1983)}]{Haldane}%
  \BibitemOpen
  \bibfield  {author} {\bibinfo {author} {\bibfnamefont {F.~D.~M.}\ \bibnamefont {Haldane}},\ }\bibfield  {title} {\bibinfo {title} {{Continuum dynamics of the 1-D Heisenberg antiferromagnet: Identification with the O(3) nonlinear sigma model}},\ }\href {https://doi.org/10.1016/0375-9601(83)90631-X} {\bibfield  {journal} {\bibinfo  {journal} {Phys. Rev. A}\ }\textbf {\bibinfo {volume} {93}},\ \bibinfo {pages} {464} (\bibinfo {year} {1983})}\BibitemShut {NoStop}%
\bibitem [{\citenamefont {Perez-Garcia}\ \emph {et~al.}(2007)\citenamefont {Perez-Garcia}, \citenamefont {Verstraete}, \citenamefont {Wolf},\ and\ \citenamefont {Cirac}}]{MPS}%
  \BibitemOpen
  \bibfield  {author} {\bibinfo {author} {\bibfnamefont {D.}~\bibnamefont {Perez-Garcia}}, \bibinfo {author} {\bibfnamefont {F.}~\bibnamefont {Verstraete}}, \bibinfo {author} {\bibfnamefont {M.~M.}\ \bibnamefont {Wolf}},\ and\ \bibinfo {author} {\bibfnamefont {J.~I.}\ \bibnamefont {Cirac}},\ }\href {https://arxiv.org/abs/quant-ph/0608197} {\bibinfo {title} {Matrix product state representations}} (\bibinfo {year} {2007}),\ \Eprint {https://arxiv.org/abs/quant-ph/0608197} {arXiv:quant-ph/0608197 [quant-ph]} \BibitemShut {NoStop}%
\bibitem [{\citenamefont {Moudgalya}\ \emph {et~al.}(2018)\citenamefont {Moudgalya}, \citenamefont {Regnault},\ and\ \citenamefont {Bernevig}}]{MPSofAKLT2}%
  \BibitemOpen
  \bibfield  {author} {\bibinfo {author} {\bibfnamefont {S.}~\bibnamefont {Moudgalya}}, \bibinfo {author} {\bibfnamefont {N.}~\bibnamefont {Regnault}},\ and\ \bibinfo {author} {\bibfnamefont {B.~A.}\ \bibnamefont {Bernevig}},\ }\bibfield  {title} {\bibinfo {title} {{Entanglement of exact excited states of Affleck-Kennedy-Lieb-Tasaki models: Exact results, many-body scars, and violation of the strong eigenstate thermalization hypothesis}},\ }\href {https://doi.org/10.1103/PhysRevB.98.235156} {\bibfield  {journal} {\bibinfo  {journal} {Phys. Rev. B}\ }\textbf {\bibinfo {volume} {98}},\ \bibinfo {pages} {235156} (\bibinfo {year} {2018})}\BibitemShut {NoStop}%
\bibitem [{\citenamefont {{\"O}stlund}\ and\ \citenamefont {Rommer}(1995)}]{MPSofAKLT3}%
  \BibitemOpen
  \bibfield  {author} {\bibinfo {author} {\bibfnamefont {S.}~\bibnamefont {{\"O}stlund}}\ and\ \bibinfo {author} {\bibfnamefont {S.}~\bibnamefont {Rommer}},\ }\bibfield  {title} {\bibinfo {title} {{Thermodynamic Limit of Density Matrix Renormalization}},\ }\href {https://doi.org/10.1103/PhysRevLett.75.3537} {\bibfield  {journal} {\bibinfo  {journal} {Phys. Rev. Lett.}\ }\textbf {\bibinfo {volume} {75}},\ \bibinfo {pages} {3537} (\bibinfo {year} {1995})}\BibitemShut {NoStop}%
\bibitem [{\citenamefont {Tasaki}(2020)}]{book}%
  \BibitemOpen
  \bibfield  {author} {\bibinfo {author} {\bibfnamefont {H.}~\bibnamefont {Tasaki}},\ }\href {https://books.google.com.hk/books?id=IU7iDwAAQBAJ} {\emph {\bibinfo {title} {{Physics and Mathematics of Quantum Many-Body Systems}}}},\ Graduate Texts in Physics\ (\bibinfo  {publisher} {{Springer International Publishing}},\ \bibinfo {year} {2020})\BibitemShut {NoStop}%
\bibitem [{\citenamefont {Affleck}\ \emph {et~al.}(1988)\citenamefont {Affleck}, \citenamefont {Kennedy}, \citenamefont {Lieb},\ and\ \citenamefont {Tasaki}}]{VBS1}%
  \BibitemOpen
  \bibfield  {author} {\bibinfo {author} {\bibfnamefont {I.}~\bibnamefont {Affleck}}, \bibinfo {author} {\bibfnamefont {T.}~\bibnamefont {Kennedy}}, \bibinfo {author} {\bibfnamefont {E.~H.}\ \bibnamefont {Lieb}},\ and\ \bibinfo {author} {\bibfnamefont {H.}~\bibnamefont {Tasaki}},\ }\bibfield  {title} {\bibinfo {title} {{Valence bond ground states in isotropic quantum antiferromagnets}},\ }\href {https://doi.org/10.1007/BF01218021} {\bibfield  {journal} {\bibinfo  {journal} {Communications in Mathematical Physics}\ }\textbf {\bibinfo {volume} {115}},\ \bibinfo {pages} {477} (\bibinfo {year} {1988})}\BibitemShut {NoStop}%
\bibitem [{\citenamefont {Arovas}\ \emph {et~al.}(1988{\natexlab{a}})\citenamefont {Arovas}, \citenamefont {Auerbach},\ and\ \citenamefont {Haldane}}]{VBS2}%
  \BibitemOpen
  \bibfield  {author} {\bibinfo {author} {\bibfnamefont {D.~P.}\ \bibnamefont {Arovas}}, \bibinfo {author} {\bibfnamefont {A.}~\bibnamefont {Auerbach}},\ and\ \bibinfo {author} {\bibfnamefont {F.~D.~M.}\ \bibnamefont {Haldane}},\ }\bibfield  {title} {\bibinfo {title} {{Extended Heisenberg models of antiferromagnetism: Analogies to the fractional quantum Hall effect}},\ }\href {https://link.aps.org/doi/10.1103/PhysRevLett.60.531} {\bibfield  {journal} {\bibinfo  {journal} {Phys. Rev. Lett.}\ }\textbf {\bibinfo {volume} {60}},\ \bibinfo {pages} {531} (\bibinfo {year} {1988}{\natexlab{a}})}\BibitemShut {NoStop}%
\bibitem [{\citenamefont {Fannes}\ \emph {et~al.}(1992{\natexlab{a}})\citenamefont {Fannes}, \citenamefont {Nachtergaele},\ and\ \citenamefont {Werner}}]{VBS3}%
  \BibitemOpen
  \bibfield  {author} {\bibinfo {author} {\bibfnamefont {M.}~\bibnamefont {Fannes}}, \bibinfo {author} {\bibfnamefont {B.}~\bibnamefont {Nachtergaele}},\ and\ \bibinfo {author} {\bibfnamefont {R.~F.}\ \bibnamefont {Werner}},\ }\bibfield  {title} {\bibinfo {title} {{Finitely correlated states on quantum spin chains}},\ }\href {https://doi.org/10.1007/BF02099178} {\bibfield  {journal} {\bibinfo  {journal} {Commun. Math. Phys.}\ }\textbf {\bibinfo {volume} {144}},\ \bibinfo {pages} {443} (\bibinfo {year} {1992}{\natexlab{a}})}\BibitemShut {NoStop}%
\bibitem [{\citenamefont {Arovas}\ \emph {et~al.}(1988{\natexlab{b}})\citenamefont {Arovas}, \citenamefont {Auerbach},\ and\ \citenamefont {Haldane}}]{highspin}%
  \BibitemOpen
  \bibfield  {author} {\bibinfo {author} {\bibfnamefont {D.~P.}\ \bibnamefont {Arovas}}, \bibinfo {author} {\bibfnamefont {A.}~\bibnamefont {Auerbach}},\ and\ \bibinfo {author} {\bibfnamefont {F.~D.~M.}\ \bibnamefont {Haldane}},\ }\bibfield  {title} {\bibinfo {title} {Extended heisenberg models of antiferromagnetism: Analogies to the fractional quantum hall effect},\ }\href {https://doi.org/10.1103/PhysRevLett.60.531} {\bibfield  {journal} {\bibinfo  {journal} {Phys. Rev. Lett.}\ }\textbf {\bibinfo {volume} {60}},\ \bibinfo {pages} {531} (\bibinfo {year} {1988}{\natexlab{b}})}\BibitemShut {NoStop}%
\bibitem [{\citenamefont {den Nijs}\ and\ \citenamefont {Rommelse}(1989)}]{SO1}%
  \BibitemOpen
  \bibfield  {author} {\bibinfo {author} {\bibfnamefont {M.}~\bibnamefont {den Nijs}}\ and\ \bibinfo {author} {\bibfnamefont {K.}~\bibnamefont {Rommelse}},\ }\bibfield  {title} {\bibinfo {title} {{Preroughening transitions in crystal surfaces and valence-bond phases in quantum spin chains}},\ }\href {https://doi.org/10.1103/PhysRevB.40.4709} {\bibfield  {journal} {\bibinfo  {journal} {Phys. Rev. B}\ }\textbf {\bibinfo {volume} {40}},\ \bibinfo {pages} {4709} (\bibinfo {year} {1989})}\BibitemShut {NoStop}%
\bibitem [{\citenamefont {P{e}rez-Garc{i}a}\ \emph {et~al.}(2008)\citenamefont {P{e}rez-Garc{i}a}, \citenamefont {Wolf}, \citenamefont {Sanz}, \citenamefont {Verstraete},\ and\ \citenamefont {Cirac}}]{SO2}%
  \BibitemOpen
  \bibfield  {author} {\bibinfo {author} {\bibfnamefont {D.}~\bibnamefont {P{e}rez-Garc{i}a}}, \bibinfo {author} {\bibfnamefont {M.~M.}\ \bibnamefont {Wolf}}, \bibinfo {author} {\bibfnamefont {M.}~\bibnamefont {Sanz}}, \bibinfo {author} {\bibfnamefont {F.}~\bibnamefont {Verstraete}},\ and\ \bibinfo {author} {\bibfnamefont {J.~I.}\ \bibnamefont {Cirac}},\ }\bibfield  {title} {\bibinfo {title} {String order and symmetries in quantum spin lattices},\ }\href {https://doi.org/10.1103/PhysRevLett.100.167202} {\bibfield  {journal} {\bibinfo  {journal} {Phys. Rev. Lett.}\ }\textbf {\bibinfo {volume} {100}},\ \bibinfo {pages} {167202} (\bibinfo {year} {2008})}\BibitemShut {NoStop}%
\bibitem [{\citenamefont {Girvin}\ and\ \citenamefont {Arovas}(1989)}]{SO3}%
  \BibitemOpen
  \bibfield  {author} {\bibinfo {author} {\bibfnamefont {S.~M.~G.}\ \bibnamefont {Girvin}}\ and\ \bibinfo {author} {\bibfnamefont {D.~P.~A.}\ \bibnamefont {Arovas}},\ }\bibfield  {title} {\bibinfo {title} {{Hidden topological order in integer quantum spin chains}},\ }\href {https://doi.org/10.1088/0031-8949/1989/T27/027} {\bibfield  {journal} {\bibinfo  {journal} {Phys. Scr.}\ }\textbf {\bibinfo {volume} {1989}},\ \bibinfo {pages} {156} (\bibinfo {year} {1989})}\BibitemShut {NoStop}%
\bibitem [{\citenamefont {Wang}\ and\ \citenamefont {Ye}(2014)}]{SO5}%
  \BibitemOpen
  \bibfield  {author} {\bibinfo {author} {\bibfnamefont {Q.-R.}\ \bibnamefont {Wang}}\ and\ \bibinfo {author} {\bibfnamefont {P.}~\bibnamefont {Ye}},\ }\bibfield  {title} {\bibinfo {title} {Sign structure and ground-state properties for a spin-s t-j chain},\ }\href {https://doi.org/10.1103/PhysRevB.90.045106} {\bibfield  {journal} {\bibinfo  {journal} {Phys. Rev. B}\ }\textbf {\bibinfo {volume} {90}},\ \bibinfo {pages} {045106} (\bibinfo {year} {2014})}\BibitemShut {NoStop}%
\bibitem [{\citenamefont {Qin}\ \emph {et~al.}(1995)\citenamefont {Qin}, \citenamefont {Ng},\ and\ \citenamefont {Su}}]{Edge}%
  \BibitemOpen
  \bibfield  {author} {\bibinfo {author} {\bibfnamefont {S.}~\bibnamefont {Qin}}, \bibinfo {author} {\bibfnamefont {T.-K.}\ \bibnamefont {Ng}},\ and\ \bibinfo {author} {\bibfnamefont {Z.-B.}\ \bibnamefont {Su}},\ }\bibfield  {title} {\bibinfo {title} {Edge states in open antiferromagnetic heisenberg chains},\ }\href {https://doi.org/10.1103/PhysRevB.52.12844} {\bibfield  {journal} {\bibinfo  {journal} {Phys. Rev. B}\ }\textbf {\bibinfo {volume} {52}},\ \bibinfo {pages} {12844} (\bibinfo {year} {1995})}\BibitemShut {NoStop}%
\bibitem [{\citenamefont {Else}\ \emph {et~al.}(2013)\citenamefont {Else}, \citenamefont {Bartlett},\ and\ \citenamefont {Doherty}}]{HS1}%
  \BibitemOpen
  \bibfield  {author} {\bibinfo {author} {\bibfnamefont {D.~V.}\ \bibnamefont {Else}}, \bibinfo {author} {\bibfnamefont {S.~D.}\ \bibnamefont {Bartlett}},\ and\ \bibinfo {author} {\bibfnamefont {A.~C.}\ \bibnamefont {Doherty}},\ }\bibfield  {title} {\bibinfo {title} {Hidden symmetry-breaking picture of symmetry-protected topological order},\ }\href {https://doi.org/10.1103/PhysRevB.88.085114} {\bibfield  {journal} {\bibinfo  {journal} {Phys. Rev. B}\ }\textbf {\bibinfo {volume} {88}},\ \bibinfo {pages} {085114} (\bibinfo {year} {2013})}\BibitemShut {NoStop}%
\bibitem [{\citenamefont {Kennedy}\ and\ \citenamefont {Tasaki}(1992{\natexlab{a}})}]{HS2}%
  \BibitemOpen
  \bibfield  {author} {\bibinfo {author} {\bibfnamefont {T.}~\bibnamefont {Kennedy}}\ and\ \bibinfo {author} {\bibfnamefont {H.}~\bibnamefont {Tasaki}},\ }\bibfield  {title} {\bibinfo {title} {{Hidden symmetry breaking and the Haldane phase in {$S=1$ }quantum spin chains}},\ }\href {https://doi.org/10.1007/BF02097239} {\bibfield  {journal} {\bibinfo  {journal} {Commun. Math. Phys.}\ }\textbf {\bibinfo {volume} {147}},\ \bibinfo {pages} {431} (\bibinfo {year} {1992}{\natexlab{a}})}\BibitemShut {NoStop}%
\bibitem [{\citenamefont {Kennedy}\ and\ \citenamefont {Tasaki}(1992{\natexlab{b}})}]{HS3}%
  \BibitemOpen
  \bibfield  {author} {\bibinfo {author} {\bibfnamefont {T.}~\bibnamefont {Kennedy}}\ and\ \bibinfo {author} {\bibfnamefont {H.}~\bibnamefont {Tasaki}},\ }\bibfield  {title} {\bibinfo {title} {{Hidden {$Z_2*Z_2$} symmetry breaking in Haldane-gap antiferromagnets}},\ }\href {https://doi.org/10.1103/PhysRevB.45.304} {\bibfield  {journal} {\bibinfo  {journal} {Phys. Rev. B}\ }\textbf {\bibinfo {volume} {45}},\ \bibinfo {pages} {304} (\bibinfo {year} {1992}{\natexlab{b}})}\BibitemShut {NoStop}%
\bibitem [{\citenamefont {Oshikawa}(1992)}]{HS4}%
  \BibitemOpen
  \bibfield  {author} {\bibinfo {author} {\bibfnamefont {M.}~\bibnamefont {Oshikawa}},\ }\bibfield  {title} {\bibinfo {title} {Hidden {$Z_2*Z_2$} symmetry in quantum spin chains with arbitrary integer spin},\ }\href {https://doi.org/10.1088/0953-8984/4/36/019} {\bibfield  {journal} {\bibinfo  {journal} {J. Phys.: Condens. Matter}\ }\textbf {\bibinfo {volume} {4}},\ \bibinfo {pages} {7469} (\bibinfo {year} {1992})}\BibitemShut {NoStop}%
\bibitem [{\citenamefont {Pollmann}\ \emph {et~al.}(2010)\citenamefont {Pollmann}, \citenamefont {Turner}, \citenamefont {Berg},\ and\ \citenamefont {Oshikawa}}]{Pollmann}%
  \BibitemOpen
  \bibfield  {author} {\bibinfo {author} {\bibfnamefont {F.}~\bibnamefont {Pollmann}}, \bibinfo {author} {\bibfnamefont {A.~M.}\ \bibnamefont {Turner}}, \bibinfo {author} {\bibfnamefont {E.}~\bibnamefont {Berg}},\ and\ \bibinfo {author} {\bibfnamefont {M.}~\bibnamefont {Oshikawa}},\ }\bibfield  {title} {\bibinfo {title} {Entanglement spectrum of a topological phase in one dimension},\ }\href {https://doi.org/10.1103/PhysRevB.81.064439} {\bibfield  {journal} {\bibinfo  {journal} {Phys. Rev. B}\ }\textbf {\bibinfo {volume} {81}},\ \bibinfo {pages} {064439} (\bibinfo {year} {2010})}\BibitemShut {NoStop}%
\bibitem [{\citenamefont {Chen}\ \emph {et~al.}(2013)\citenamefont {Chen}, \citenamefont {Gu}, \citenamefont {Liu},\ and\ \citenamefont {Wen}}]{Chen2013CGLWbSPT}%
  \BibitemOpen
  \bibfield  {author} {\bibinfo {author} {\bibfnamefont {X.}~\bibnamefont {Chen}}, \bibinfo {author} {\bibfnamefont {Z.-C.}\ \bibnamefont {Gu}}, \bibinfo {author} {\bibfnamefont {Z.-X.}\ \bibnamefont {Liu}},\ and\ \bibinfo {author} {\bibfnamefont {X.-G.}\ \bibnamefont {Wen}},\ }\bibfield  {title} {\bibinfo {title} {Symmetry protected topological orders and the group cohomology of their symmetry group},\ }\href {https://doi.org/10.1103/PhysRevB.87.155114} {\bibfield  {journal} {\bibinfo  {journal} {Phys. Rev. B}\ }\textbf {\bibinfo {volume} {87}},\ \bibinfo {pages} {155114} (\bibinfo {year} {2013})}\BibitemShut {NoStop}%
\bibitem [{\citenamefont {Kapustin}()}]{Kapustin2014Cobordism}%
  \BibitemOpen
  \bibfield  {author} {\bibinfo {author} {\bibfnamefont {A.}~\bibnamefont {Kapustin}},\ }\bibfield  {title} {\bibinfo {title} {Symmetry protected topological phases, anomalies, and cobordisms: Beyond group cohomology},\ }\href {https://arxiv.org/abs/1403.1467} {\ }\Eprint {https://arxiv.org/abs/arXiv:1403.1467} {arXiv:1403.1467} \BibitemShut {NoStop}%
\bibitem [{\citenamefont {Kapustin}\ \emph {et~al.}(2015)\citenamefont {Kapustin}, \citenamefont {Thorngren}, \citenamefont {Turzillo},\ and\ \citenamefont {Wang}}]{Kapustin2015fCobordism}%
  \BibitemOpen
  \bibfield  {author} {\bibinfo {author} {\bibfnamefont {A.}~\bibnamefont {Kapustin}}, \bibinfo {author} {\bibfnamefont {R.}~\bibnamefont {Thorngren}}, \bibinfo {author} {\bibfnamefont {A.}~\bibnamefont {Turzillo}},\ and\ \bibinfo {author} {\bibfnamefont {Z.}~\bibnamefont {Wang}},\ }\bibfield  {title} {\bibinfo {title} {Fermionic symmetry protected topological phases and cobordisms},\ }\href {https://doi.org/10.48550/arXiv.1406.7329} {\bibfield  {journal} {\bibinfo  {journal} {Journal of High Energy Physics}\ }\textbf {\bibinfo {volume} {2015}},\ \bibinfo {pages} {1} (\bibinfo {year} {2015})}\BibitemShut {NoStop}%
\bibitem [{\citenamefont {Bi}\ \emph {et~al.}(2015)\citenamefont {Bi}, \citenamefont {Rasmussen}, \citenamefont {Slagle},\ and\ \citenamefont {Xu}}]{Bi2015NLSMSPT}%
  \BibitemOpen
  \bibfield  {author} {\bibinfo {author} {\bibfnamefont {Z.}~\bibnamefont {Bi}}, \bibinfo {author} {\bibfnamefont {A.}~\bibnamefont {Rasmussen}}, \bibinfo {author} {\bibfnamefont {K.}~\bibnamefont {Slagle}},\ and\ \bibinfo {author} {\bibfnamefont {C.}~\bibnamefont {Xu}},\ }\bibfield  {title} {\bibinfo {title} {Classification and description of bosonic symmetry protected topological phases with semiclassical nonlinear sigma models},\ }\href {https://doi.org/10.1103/PhysRevB.91.134404} {\bibfield  {journal} {\bibinfo  {journal} {Phys. Rev. B}\ }\textbf {\bibinfo {volume} {91}},\ \bibinfo {pages} {134404} (\bibinfo {year} {2015})}\BibitemShut {NoStop}%
\bibitem [{\citenamefont {You}\ and\ \citenamefont {You}(2016)}]{YouYou2016bSPT}%
  \BibitemOpen
  \bibfield  {author} {\bibinfo {author} {\bibfnamefont {Y.}~\bibnamefont {You}}\ and\ \bibinfo {author} {\bibfnamefont {Y.-Z.}\ \bibnamefont {You}},\ }\bibfield  {title} {\bibinfo {title} {Geometry defects in bosonic symmetry-protected topological phases},\ }\href {https://doi.org/10.1103/PhysRevB.93.245135} {\bibfield  {journal} {\bibinfo  {journal} {Phys. Rev. B}\ }\textbf {\bibinfo {volume} {93}},\ \bibinfo {pages} {245135} (\bibinfo {year} {2016})}\BibitemShut {NoStop}%
\bibitem [{\citenamefont {Lu}\ and\ \citenamefont {Vishwanath}(2012)}]{Lu2012CSSPT}%
  \BibitemOpen
  \bibfield  {author} {\bibinfo {author} {\bibfnamefont {Y.-M.}\ \bibnamefont {Lu}}\ and\ \bibinfo {author} {\bibfnamefont {A.}~\bibnamefont {Vishwanath}},\ }\bibfield  {title} {\bibinfo {title} {Theory and classification of interacting integer topological phases in two dimensions: A chern-simons approach},\ }\href {https://doi.org/10.1103/PhysRevB.86.125119} {\bibfield  {journal} {\bibinfo  {journal} {Phys. Rev. B}\ }\textbf {\bibinfo {volume} {86}},\ \bibinfo {pages} {125119} (\bibinfo {year} {2012})}\BibitemShut {NoStop}%
\bibitem [{\citenamefont {Ye}\ and\ \citenamefont {Gu}(2015)}]{bti2}%
  \BibitemOpen
  \bibfield  {author} {\bibinfo {author} {\bibfnamefont {P.}~\bibnamefont {Ye}}\ and\ \bibinfo {author} {\bibfnamefont {Z.-C.}\ \bibnamefont {Gu}},\ }\bibfield  {title} {\bibinfo {title} {Vortex-line condensation in three dimensions: A physical mechanism for bosonic topological insulators},\ }\href {https://doi.org/10.1103/PhysRevX.5.021029} {\bibfield  {journal} {\bibinfo  {journal} {Phys. Rev. X}\ }\textbf {\bibinfo {volume} {5}},\ \bibinfo {pages} {021029} (\bibinfo {year} {2015})}\BibitemShut {NoStop}%
\bibitem [{\citenamefont {Gu}\ \emph {et~al.}(2016)\citenamefont {Gu}, \citenamefont {Wang},\ and\ \citenamefont {Wen}}]{Gu2016SPTBF}%
  \BibitemOpen
  \bibfield  {author} {\bibinfo {author} {\bibfnamefont {Z.-C.}\ \bibnamefont {Gu}}, \bibinfo {author} {\bibfnamefont {J.~C.}\ \bibnamefont {Wang}},\ and\ \bibinfo {author} {\bibfnamefont {X.-G.}\ \bibnamefont {Wen}},\ }\bibfield  {title} {\bibinfo {title} {Multikink topological terms and charge-binding domain-wall condensation induced symmetry-protected topological states: Beyond chern-simons/bf field theories},\ }\href {https://doi.org/10.1103/PhysRevB.93.115136} {\bibfield  {journal} {\bibinfo  {journal} {Phys. Rev. B}\ }\textbf {\bibinfo {volume} {93}},\ \bibinfo {pages} {115136} (\bibinfo {year} {2016})}\BibitemShut {NoStop}%
\bibitem [{\citenamefont {Ye}\ and\ \citenamefont {Gu}(2016)}]{Ye20163dbSPT}%
  \BibitemOpen
  \bibfield  {author} {\bibinfo {author} {\bibfnamefont {P.}~\bibnamefont {Ye}}\ and\ \bibinfo {author} {\bibfnamefont {Z.-C.}\ \bibnamefont {Gu}},\ }\bibfield  {title} {\bibinfo {title} {Topological quantum field theory of three-dimensional bosonic abelian-symmetry-protected topological phases},\ }\href {https://doi.org/10.1103/PhysRevB.93.205157} {\bibfield  {journal} {\bibinfo  {journal} {Phys. Rev. B}\ }\textbf {\bibinfo {volume} {93}},\ \bibinfo {pages} {205157} (\bibinfo {year} {2016})}\BibitemShut {NoStop}%
\bibitem [{\citenamefont {Wang}\ \emph {et~al.}(2018)\citenamefont {Wang}, \citenamefont {Ohmori}, \citenamefont {Putrov}, \citenamefont {Zheng}, \citenamefont {Wan}, \citenamefont {Guo}, \citenamefont {Lin}, \citenamefont {Gao},\ and\ \citenamefont {Yau}}]{2018arXiv180105416W}%
  \BibitemOpen
  \bibfield  {author} {\bibinfo {author} {\bibfnamefont {J.}~\bibnamefont {Wang}}, \bibinfo {author} {\bibfnamefont {K.}~\bibnamefont {Ohmori}}, \bibinfo {author} {\bibfnamefont {P.}~\bibnamefont {Putrov}}, \bibinfo {author} {\bibfnamefont {Y.}~\bibnamefont {Zheng}}, \bibinfo {author} {\bibfnamefont {Z.}~\bibnamefont {Wan}}, \bibinfo {author} {\bibfnamefont {M.}~\bibnamefont {Guo}}, \bibinfo {author} {\bibfnamefont {H.}~\bibnamefont {Lin}}, \bibinfo {author} {\bibfnamefont {P.}~\bibnamefont {Gao}},\ and\ \bibinfo {author} {\bibfnamefont {S.-T.}\ \bibnamefont {Yau}},\ }\bibfield  {title} {\bibinfo {title} {Tunneling topological vacua via extended operators: (spin-)tqft spectra and boundary deconfinement in various dimensions},\ }\href {https://doi.org/10.1093/ptep/pty051} {\bibfield  {journal} {\bibinfo  {journal} {Progress of Theoretical and Experimental Physics}\ }\textbf {\bibinfo {volume} {2018}},\ \bibinfo {pages} {053A01} (\bibinfo {year} {2018})}\BibitemShut {NoStop}%
\bibitem [{\citenamefont {Hsieh}\ \emph {et~al.}(2014)\citenamefont {Hsieh}, \citenamefont {Sule}, \citenamefont {Cho}, \citenamefont {Ryu},\ and\ \citenamefont {Leigh}}]{Hsieh2014SPT2dorbifold}%
  \BibitemOpen
  \bibfield  {author} {\bibinfo {author} {\bibfnamefont {C.-T.}\ \bibnamefont {Hsieh}}, \bibinfo {author} {\bibfnamefont {O.~M.}\ \bibnamefont {Sule}}, \bibinfo {author} {\bibfnamefont {G.~Y.}\ \bibnamefont {Cho}}, \bibinfo {author} {\bibfnamefont {S.}~\bibnamefont {Ryu}},\ and\ \bibinfo {author} {\bibfnamefont {R.~G.}\ \bibnamefont {Leigh}},\ }\bibfield  {title} {\bibinfo {title} {Symmetry-protected topological phases, generalized laughlin argument, and orientifolds},\ }\href {https://doi.org/10.1103/PhysRevB.90.165134} {\bibfield  {journal} {\bibinfo  {journal} {Phys. Rev. B}\ }\textbf {\bibinfo {volume} {90}},\ \bibinfo {pages} {165134} (\bibinfo {year} {2014})}\BibitemShut {NoStop}%
\bibitem [{\citenamefont {Hsieh}\ \emph {et~al.}(2016)\citenamefont {Hsieh}, \citenamefont {Cho},\ and\ \citenamefont {Ryu}}]{Hsieh2016SPT3dorbifold}%
  \BibitemOpen
  \bibfield  {author} {\bibinfo {author} {\bibfnamefont {C.-T.}\ \bibnamefont {Hsieh}}, \bibinfo {author} {\bibfnamefont {G.~Y.}\ \bibnamefont {Cho}},\ and\ \bibinfo {author} {\bibfnamefont {S.}~\bibnamefont {Ryu}},\ }\bibfield  {title} {\bibinfo {title} {Global anomalies on the surface of fermionic symmetry-protected topological phases in (3+1) dimensions},\ }\href {https://doi.org/10.1103/PhysRevB.93.075135} {\bibfield  {journal} {\bibinfo  {journal} {Phys. Rev. B}\ }\textbf {\bibinfo {volume} {93}},\ \bibinfo {pages} {075135} (\bibinfo {year} {2016})}\BibitemShut {NoStop}%
\bibitem [{\citenamefont {Han}\ \emph {et~al.}(2017)\citenamefont {Han}, \citenamefont {Tiwari}, \citenamefont {Hsieh},\ and\ \citenamefont {Ryu}}]{Han2017BCFTSPT}%
  \BibitemOpen
  \bibfield  {author} {\bibinfo {author} {\bibfnamefont {B.}~\bibnamefont {Han}}, \bibinfo {author} {\bibfnamefont {A.}~\bibnamefont {Tiwari}}, \bibinfo {author} {\bibfnamefont {C.-T.}\ \bibnamefont {Hsieh}},\ and\ \bibinfo {author} {\bibfnamefont {S.}~\bibnamefont {Ryu}},\ }\bibfield  {title} {\bibinfo {title} {Boundary conformal field theory and symmetry-protected topological phases in $2+1$ dimensions},\ }\href {https://doi.org/10.1103/PhysRevB.96.125105} {\bibfield  {journal} {\bibinfo  {journal} {Phys. Rev. B}\ }\textbf {\bibinfo {volume} {96}},\ \bibinfo {pages} {125105} (\bibinfo {year} {2017})}\BibitemShut {NoStop}%
\bibitem [{\citenamefont {Chen}\ \emph {et~al.}(2014)\citenamefont {Chen}, \citenamefont {Lu},\ and\ \citenamefont {Vishwanath}}]{Chen:2014aa}%
  \BibitemOpen
  \bibfield  {author} {\bibinfo {author} {\bibfnamefont {X.}~\bibnamefont {Chen}}, \bibinfo {author} {\bibfnamefont {Y.-M.}\ \bibnamefont {Lu}},\ and\ \bibinfo {author} {\bibfnamefont {A.}~\bibnamefont {Vishwanath}},\ }\bibfield  {title} {\bibinfo {title} {Symmetry-protected topological phases from decorated domain walls},\ }\href {http://dx.doi.org/10.1038/ncomms4507} {\bibfield  {journal} {\bibinfo  {journal} {Nature Communications}\ }\textbf {\bibinfo {volume} {5}},\ \bibinfo {pages} {3507 EP } (\bibinfo {year} {2014})}\BibitemShut {NoStop}%
\bibitem [{\citenamefont {Ye}\ and\ \citenamefont {Wen}(2013)}]{YW12}%
  \BibitemOpen
  \bibfield  {author} {\bibinfo {author} {\bibfnamefont {P.}~\bibnamefont {Ye}}\ and\ \bibinfo {author} {\bibfnamefont {X.-G.}\ \bibnamefont {Wen}},\ }\bibfield  {title} {\bibinfo {title} {Projective construction of two-dimensional symmetry-protected topological phases with u(1), so(3), or su(2) symmetries},\ }\href {https://doi.org/10.1103/PhysRevB.87.195128} {\bibfield  {journal} {\bibinfo  {journal} {Phys. Rev. B}\ }\textbf {\bibinfo {volume} {87}},\ \bibinfo {pages} {195128} (\bibinfo {year} {2013})}\BibitemShut {NoStop}%
\bibitem [{\citenamefont {Lu}\ and\ \citenamefont {Lee}(2014)}]{LL1263}%
  \BibitemOpen
  \bibfield  {author} {\bibinfo {author} {\bibfnamefont {Y.-M.}\ \bibnamefont {Lu}}\ and\ \bibinfo {author} {\bibfnamefont {D.-H.}\ \bibnamefont {Lee}},\ }\bibfield  {title} {\bibinfo {title} {Spin quantum hall effects in featureless nonfractionalized spin-1 magnets},\ }\href {https://doi.org/10.1103/PhysRevB.89.184417} {\bibfield  {journal} {\bibinfo  {journal} {Phys. Rev. B}\ }\textbf {\bibinfo {volume} {89}},\ \bibinfo {pages} {184417} (\bibinfo {year} {2014})}\BibitemShut {NoStop}%
\bibitem [{\citenamefont {Liu}\ \emph {et~al.}(2014)\citenamefont {Liu}, \citenamefont {Mei}, \citenamefont {Ye},\ and\ \citenamefont {Wen}}]{Ye14b}%
  \BibitemOpen
  \bibfield  {author} {\bibinfo {author} {\bibfnamefont {Z.-X.}\ \bibnamefont {Liu}}, \bibinfo {author} {\bibfnamefont {J.-W.}\ \bibnamefont {Mei}}, \bibinfo {author} {\bibfnamefont {P.}~\bibnamefont {Ye}},\ and\ \bibinfo {author} {\bibfnamefont {X.-G.}\ \bibnamefont {Wen}},\ }\bibfield  {title} {\bibinfo {title} {$u(1)\ifmmode\times\else\texttimes\fi{}u(1)$ symmetry-protected topological order in gutzwiller wave functions},\ }\href {https://doi.org/10.1103/PhysRevB.90.235146} {\bibfield  {journal} {\bibinfo  {journal} {Phys. Rev. B}\ }\textbf {\bibinfo {volume} {90}},\ \bibinfo {pages} {235146} (\bibinfo {year} {2014})}\BibitemShut {NoStop}%
\bibitem [{\citenamefont {Ye}\ and\ \citenamefont {Wen}(2014)}]{YW13a}%
  \BibitemOpen
  \bibfield  {author} {\bibinfo {author} {\bibfnamefont {P.}~\bibnamefont {Ye}}\ and\ \bibinfo {author} {\bibfnamefont {X.-G.}\ \bibnamefont {Wen}},\ }\bibfield  {title} {\bibinfo {title} {Constructing symmetric topological phases of bosons in three dimensions via fermionic projective construction and dyon condensation},\ }\href {https://doi.org/10.1103/PhysRevB.89.045127} {\bibfield  {journal} {\bibinfo  {journal} {Phys. Rev. B}\ }\textbf {\bibinfo {volume} {89}},\ \bibinfo {pages} {045127} (\bibinfo {year} {2014})}\BibitemShut {NoStop}%
\bibitem [{\citenamefont {Ye}\ \emph {et~al.}(2016)\citenamefont {Ye}, \citenamefont {Hughes}, \citenamefont {Maciejko},\ and\ \citenamefont {Fradkin}}]{ye16a}%
  \BibitemOpen
  \bibfield  {author} {\bibinfo {author} {\bibfnamefont {P.}~\bibnamefont {Ye}}, \bibinfo {author} {\bibfnamefont {T.~L.}\ \bibnamefont {Hughes}}, \bibinfo {author} {\bibfnamefont {J.}~\bibnamefont {Maciejko}},\ and\ \bibinfo {author} {\bibfnamefont {E.}~\bibnamefont {Fradkin}},\ }\bibfield  {title} {\bibinfo {title} {Composite particle theory of three-dimensional gapped fermionic phases: Fractional topological insulators and charge-loop excitation symmetry},\ }\href {https://doi.org/10.1103/PhysRevB.94.115104} {\bibfield  {journal} {\bibinfo  {journal} {Phys. Rev. B}\ }\textbf {\bibinfo {volume} {94}},\ \bibinfo {pages} {115104} (\bibinfo {year} {2016})}\BibitemShut {NoStop}%
\bibitem [{\citenamefont {Wang}\ \emph {et~al.}(2015{\natexlab{a}})\citenamefont {Wang}, \citenamefont {Nahum},\ and\ \citenamefont {Senthil}}]{Wangsenthil2015}%
  \BibitemOpen
  \bibfield  {author} {\bibinfo {author} {\bibfnamefont {C.}~\bibnamefont {Wang}}, \bibinfo {author} {\bibfnamefont {A.}~\bibnamefont {Nahum}},\ and\ \bibinfo {author} {\bibfnamefont {T.}~\bibnamefont {Senthil}},\ }\bibfield  {title} {\bibinfo {title} {Topological paramagnetism in frustrated spin-1 mott insulators},\ }\href {https://doi.org/10.1103/PhysRevB.91.195131} {\bibfield  {journal} {\bibinfo  {journal} {Phys. Rev. B}\ }\textbf {\bibinfo {volume} {91}},\ \bibinfo {pages} {195131} (\bibinfo {year} {2015}{\natexlab{a}})}\BibitemShut {NoStop}%
\bibitem [{\citenamefont {Cheng}\ and\ \citenamefont {Gu}(2014)}]{PhysRevLett.112.141602}%
  \BibitemOpen
  \bibfield  {author} {\bibinfo {author} {\bibfnamefont {M.}~\bibnamefont {Cheng}}\ and\ \bibinfo {author} {\bibfnamefont {Z.-C.}\ \bibnamefont {Gu}},\ }\bibfield  {title} {\bibinfo {title} {Topological response theory of abelian symmetry-protected topological phases in two dimensions},\ }\href {https://doi.org/10.1103/PhysRevLett.112.141602} {\bibfield  {journal} {\bibinfo  {journal} {Phys. Rev. Lett.}\ }\textbf {\bibinfo {volume} {112}},\ \bibinfo {pages} {141602} (\bibinfo {year} {2014})}\BibitemShut {NoStop}%
\bibitem [{\citenamefont {Hung}\ and\ \citenamefont {Wen}(2013)}]{Hung_Wen_gauge}%
  \BibitemOpen
  \bibfield  {author} {\bibinfo {author} {\bibfnamefont {L.-Y.}\ \bibnamefont {Hung}}\ and\ \bibinfo {author} {\bibfnamefont {X.-G.}\ \bibnamefont {Wen}},\ }\bibfield  {title} {\bibinfo {title} {Quantized topological terms in weak-coupling gauge theories with a global symmetry and their connection to symmetry-enriched topological phases},\ }\href {https://doi.org/10.1103/PhysRevB.87.165107} {\bibfield  {journal} {\bibinfo  {journal} {Phys. Rev. B}\ }\textbf {\bibinfo {volume} {87}},\ \bibinfo {pages} {165107} (\bibinfo {year} {2013})}\BibitemShut {NoStop}%
\bibitem [{\citenamefont {Wen}(2013)}]{PhysRevD.88.045013}%
  \BibitemOpen
  \bibfield  {author} {\bibinfo {author} {\bibfnamefont {X.-G.}\ \bibnamefont {Wen}},\ }\bibfield  {title} {\bibinfo {title} {Classifying gauge anomalies through symmetry-protected trivial orders and classifying gravitational anomalies through topological orders},\ }\href {https://doi.org/10.1103/PhysRevD.88.045013} {\bibfield  {journal} {\bibinfo  {journal} {Phys. Rev. D}\ }\textbf {\bibinfo {volume} {88}},\ \bibinfo {pages} {045013} (\bibinfo {year} {2013})}\BibitemShut {NoStop}%
\bibitem [{\citenamefont {Ye}\ and\ \citenamefont {Wang}(2013)}]{bti6}%
  \BibitemOpen
  \bibfield  {author} {\bibinfo {author} {\bibfnamefont {P.}~\bibnamefont {Ye}}\ and\ \bibinfo {author} {\bibfnamefont {J.}~\bibnamefont {Wang}},\ }\bibfield  {title} {\bibinfo {title} {Symmetry-protected topological phases with charge and spin symmetries: Response theory and dynamical gauge theory in two and three dimensions},\ }\href {https://doi.org/10.1103/PhysRevB.88.235109} {\bibfield  {journal} {\bibinfo  {journal} {Phys. Rev. B}\ }\textbf {\bibinfo {volume} {88}},\ \bibinfo {pages} {235109} (\bibinfo {year} {2013})}\BibitemShut {NoStop}%
\bibitem [{\citenamefont {Lapa}\ \emph {et~al.}(2017)\citenamefont {Lapa}, \citenamefont {Jian}, \citenamefont {Ye},\ and\ \citenamefont {Hughes}}]{lapa17}%
  \BibitemOpen
  \bibfield  {author} {\bibinfo {author} {\bibfnamefont {M.~F.}\ \bibnamefont {Lapa}}, \bibinfo {author} {\bibfnamefont {C.-M.}\ \bibnamefont {Jian}}, \bibinfo {author} {\bibfnamefont {P.}~\bibnamefont {Ye}},\ and\ \bibinfo {author} {\bibfnamefont {T.~L.}\ \bibnamefont {Hughes}},\ }\bibfield  {title} {\bibinfo {title} {Topological electromagnetic responses of bosonic quantum hall, topological insulator, and chiral semimetal phases in all dimensions},\ }\href {https://doi.org/10.1103/PhysRevB.95.035149} {\bibfield  {journal} {\bibinfo  {journal} {Phys. Rev. B}\ }\textbf {\bibinfo {volume} {95}},\ \bibinfo {pages} {035149} (\bibinfo {year} {2017})}\BibitemShut {NoStop}%
\bibitem [{\citenamefont {Wang}\ \emph {et~al.}(2015{\natexlab{b}})\citenamefont {Wang}, \citenamefont {Gu},\ and\ \citenamefont {Wen}}]{Wang2015GaugeGravitySPT}%
  \BibitemOpen
  \bibfield  {author} {\bibinfo {author} {\bibfnamefont {J.~C.}\ \bibnamefont {Wang}}, \bibinfo {author} {\bibfnamefont {Z.-C.}\ \bibnamefont {Gu}},\ and\ \bibinfo {author} {\bibfnamefont {X.-G.}\ \bibnamefont {Wen}},\ }\bibfield  {title} {\bibinfo {title} {Field-theory representation of gauge-gravity symmetry-protected topological invariants, group cohomology, and beyond},\ }\href {https://doi.org/10.1103/PhysRevLett.114.031601} {\bibfield  {journal} {\bibinfo  {journal} {Phys. Rev. Lett.}\ }\textbf {\bibinfo {volume} {114}},\ \bibinfo {pages} {031601} (\bibinfo {year} {2015}{\natexlab{b}})}\BibitemShut {NoStop}%
\bibitem [{\citenamefont {Han}\ \emph {et~al.}(2019)\citenamefont {Han}, \citenamefont {Wang},\ and\ \citenamefont {Ye}}]{PhysRevB.99.205120}%
  \BibitemOpen
  \bibfield  {author} {\bibinfo {author} {\bibfnamefont {B.}~\bibnamefont {Han}}, \bibinfo {author} {\bibfnamefont {H.}~\bibnamefont {Wang}},\ and\ \bibinfo {author} {\bibfnamefont {P.}~\bibnamefont {Ye}},\ }\bibfield  {title} {\bibinfo {title} {Generalized wen-zee terms},\ }\href {https://doi.org/10.1103/PhysRevB.99.205120} {\bibfield  {journal} {\bibinfo  {journal} {Phys. Rev. B}\ }\textbf {\bibinfo {volume} {99}},\ \bibinfo {pages} {205120} (\bibinfo {year} {2019})}\BibitemShut {NoStop}%
\bibitem [{\citenamefont {Levin}\ and\ \citenamefont {Gu}(2012)}]{levin_gu_12}%
  \BibitemOpen
  \bibfield  {author} {\bibinfo {author} {\bibfnamefont {M.}~\bibnamefont {Levin}}\ and\ \bibinfo {author} {\bibfnamefont {Z.-C.}\ \bibnamefont {Gu}},\ }\bibfield  {title} {\bibinfo {title} {Braiding statistics approach to symmetry-protected topological phases},\ }\href {https://doi.org/10.1103/PhysRevB.86.115109} {\bibfield  {journal} {\bibinfo  {journal} {Phys. Rev. B}\ }\textbf {\bibinfo {volume} {86}},\ \bibinfo {pages} {115109} (\bibinfo {year} {2012})}\BibitemShut {NoStop}%
\bibitem [{\citenamefont {Wang}\ and\ \citenamefont {Levin}(2014)}]{wang_levin1}%
  \BibitemOpen
  \bibfield  {author} {\bibinfo {author} {\bibfnamefont {C.}~\bibnamefont {Wang}}\ and\ \bibinfo {author} {\bibfnamefont {M.}~\bibnamefont {Levin}},\ }\bibfield  {title} {\bibinfo {title} {Braiding statistics of loop excitations in three dimensions},\ }\href {https://doi.org/10.1103/PhysRevLett.113.080403} {\bibfield  {journal} {\bibinfo  {journal} {Phys. Rev. Lett.}\ }\textbf {\bibinfo {volume} {113}},\ \bibinfo {pages} {080403} (\bibinfo {year} {2014})}\BibitemShut {NoStop}%
\bibitem [{\citenamefont {Putrov}\ \emph {et~al.}(2017)\citenamefont {Putrov}, \citenamefont {Wang},\ and\ \citenamefont {Yau}}]{2016arXiv161209298P}%
  \BibitemOpen
  \bibfield  {author} {\bibinfo {author} {\bibfnamefont {P.}~\bibnamefont {Putrov}}, \bibinfo {author} {\bibfnamefont {J.}~\bibnamefont {Wang}},\ and\ \bibinfo {author} {\bibfnamefont {S.-T.}\ \bibnamefont {Yau}},\ }\bibfield  {title} {\bibinfo {title} {Braiding statistics and link invariants of bosonic/fermionic topological quantum matter in 2+1 and 3+1 dimensions},\ }\href {https://doi.org/https://doi.org/10.1016/j.aop.2017.06.019} {\bibfield  {journal} {\bibinfo  {journal} {Annals of Physics}\ }\textbf {\bibinfo {volume} {384}},\ \bibinfo {pages} {254 } (\bibinfo {year} {2017})}\BibitemShut {NoStop}%
\bibitem [{\citenamefont {Chan}\ \emph {et~al.}(2018)\citenamefont {Chan}, \citenamefont {Ye},\ and\ \citenamefont {Ryu}}]{ye17b}%
  \BibitemOpen
  \bibfield  {author} {\bibinfo {author} {\bibfnamefont {A.~P.~O.}\ \bibnamefont {Chan}}, \bibinfo {author} {\bibfnamefont {P.}~\bibnamefont {Ye}},\ and\ \bibinfo {author} {\bibfnamefont {S.}~\bibnamefont {Ryu}},\ }\bibfield  {title} {\bibinfo {title} {Braiding with borromean rings in ($3+1$)-dimensional spacetime},\ }\href {https://doi.org/10.1103/PhysRevLett.121.061601} {\bibfield  {journal} {\bibinfo  {journal} {Phys. Rev. Lett.}\ }\textbf {\bibinfo {volume} {121}},\ \bibinfo {pages} {061601} (\bibinfo {year} {2018})}\BibitemShut {NoStop}%
\bibitem [{\citenamefont {Chen}\ \emph {et~al.}(2011{\natexlab{b}})\citenamefont {Chen}, \citenamefont {Liu},\ and\ \citenamefont {Wen}}]{PhysRevB.84.235141}%
  \BibitemOpen
  \bibfield  {author} {\bibinfo {author} {\bibfnamefont {X.}~\bibnamefont {Chen}}, \bibinfo {author} {\bibfnamefont {Z.-X.}\ \bibnamefont {Liu}},\ and\ \bibinfo {author} {\bibfnamefont {X.-G.}\ \bibnamefont {Wen}},\ }\bibfield  {title} {\bibinfo {title} {Two-dimensional symmetry-protected topological orders and their protected gapless edge excitations},\ }\href {https://doi.org/10.1103/PhysRevB.84.235141} {\bibfield  {journal} {\bibinfo  {journal} {Phys. Rev. B}\ }\textbf {\bibinfo {volume} {84}},\ \bibinfo {pages} {235141} (\bibinfo {year} {2011}{\natexlab{b}})}\BibitemShut {NoStop}%
\bibitem [{\citenamefont {Chen}\ \emph {et~al.}(2012)\citenamefont {Chen}, \citenamefont {Gu}, \citenamefont {Liu},\ and\ \citenamefont {Wen}}]{Chen_science}%
  \BibitemOpen
  \bibfield  {author} {\bibinfo {author} {\bibfnamefont {X.}~\bibnamefont {Chen}}, \bibinfo {author} {\bibfnamefont {Z.-C.}\ \bibnamefont {Gu}}, \bibinfo {author} {\bibfnamefont {Z.-X.}\ \bibnamefont {Liu}},\ and\ \bibinfo {author} {\bibfnamefont {X.-G.}\ \bibnamefont {Wen}},\ }\bibfield  {title} {\bibinfo {title} {Symmetry-protected topological orders in interacting bosonic systems},\ }\href {https://doi.org/10.1126/science.1227224} {\bibfield  {journal} {\bibinfo  {journal} {Science}\ }\textbf {\bibinfo {volume} {338}},\ \bibinfo {pages} {1604} (\bibinfo {year} {2012})}\BibitemShut {NoStop}%
\bibitem [{\citenamefont {Ning}\ \emph {et~al.}(2016)\citenamefont {Ning}, \citenamefont {Liu},\ and\ \citenamefont {Ye}}]{ye16_set}%
  \BibitemOpen
  \bibfield  {author} {\bibinfo {author} {\bibfnamefont {S.-Q.}\ \bibnamefont {Ning}}, \bibinfo {author} {\bibfnamefont {Z.-X.}\ \bibnamefont {Liu}},\ and\ \bibinfo {author} {\bibfnamefont {P.}~\bibnamefont {Ye}},\ }\bibfield  {title} {\bibinfo {title} {Symmetry enrichment in three-dimensional topological phases},\ }\href {https://doi.org/10.1103/PhysRevB.94.245120} {\bibfield  {journal} {\bibinfo  {journal} {Phys. Rev. B}\ }\textbf {\bibinfo {volume} {94}},\ \bibinfo {pages} {245120} (\bibinfo {year} {2016})}\BibitemShut {NoStop}%
\bibitem [{\citenamefont {You}\ \emph {et~al.}(2014)\citenamefont {You}, \citenamefont {Bi}, \citenamefont {Rasmussen}, \citenamefont {Slagle},\ and\ \citenamefont {Xu}}]{Wave2014You}%
  \BibitemOpen
  \bibfield  {author} {\bibinfo {author} {\bibfnamefont {Y.-Z.}\ \bibnamefont {You}}, \bibinfo {author} {\bibfnamefont {Z.}~\bibnamefont {Bi}}, \bibinfo {author} {\bibfnamefont {A.}~\bibnamefont {Rasmussen}}, \bibinfo {author} {\bibfnamefont {K.}~\bibnamefont {Slagle}},\ and\ \bibinfo {author} {\bibfnamefont {C.}~\bibnamefont {Xu}},\ }\bibfield  {title} {\bibinfo {title} {Wave function and strange correlator of short-range entangled states},\ }\href {https://doi.org/10.1103/PhysRevLett.112.247202} {\bibfield  {journal} {\bibinfo  {journal} {Phys. Rev. Lett.}\ }\textbf {\bibinfo {volume} {112}},\ \bibinfo {pages} {247202} (\bibinfo {year} {2014})}\BibitemShut {NoStop}%
\bibitem [{\citenamefont {Wu}\ \emph {et~al.}(2015)\citenamefont {Wu}, \citenamefont {He}, \citenamefont {You}, \citenamefont {Xu}, \citenamefont {Meng},\ and\ \citenamefont {Lu}}]{Quantum2015Wu}%
  \BibitemOpen
  \bibfield  {author} {\bibinfo {author} {\bibfnamefont {H.-Q.}\ \bibnamefont {Wu}}, \bibinfo {author} {\bibfnamefont {Y.-Y.}\ \bibnamefont {He}}, \bibinfo {author} {\bibfnamefont {Y.-Z.}\ \bibnamefont {You}}, \bibinfo {author} {\bibfnamefont {C.}~\bibnamefont {Xu}}, \bibinfo {author} {\bibfnamefont {Z.~Y.}\ \bibnamefont {Meng}},\ and\ \bibinfo {author} {\bibfnamefont {Z.-Y.}\ \bibnamefont {Lu}},\ }\bibfield  {title} {\bibinfo {title} {Quantum monte carlo study of strange correlator in interacting topological insulators},\ }\href {https://doi.org/10.1103/PhysRevB.92.165123} {\bibfield  {journal} {\bibinfo  {journal} {Phys. Rev. B}\ }\textbf {\bibinfo {volume} {92}},\ \bibinfo {pages} {165123} (\bibinfo {year} {2015})}\BibitemShut {NoStop}%
\bibitem [{\citenamefont {Wierschem}\ and\ \citenamefont {Sengupta}(2014)}]{Strange2014Wierschem}%
  \BibitemOpen
  \bibfield  {author} {\bibinfo {author} {\bibfnamefont {K.}~\bibnamefont {Wierschem}}\ and\ \bibinfo {author} {\bibfnamefont {P.}~\bibnamefont {Sengupta}},\ }\bibfield  {title} {\bibinfo {title} {Strange correlations in spin-1 heisenberg antiferromagnets},\ }\href {https://doi.org/10.1103/PhysRevB.90.115157} {\bibfield  {journal} {\bibinfo  {journal} {Phys. Rev. B}\ }\textbf {\bibinfo {volume} {90}},\ \bibinfo {pages} {115157} (\bibinfo {year} {2014})}\BibitemShut {NoStop}%
\bibitem [{\citenamefont {Wierschem}\ and\ \citenamefont {Beach}(2016{\natexlab{a}})}]{Detection2016Wierschem}%
  \BibitemOpen
  \bibfield  {author} {\bibinfo {author} {\bibfnamefont {K.}~\bibnamefont {Wierschem}}\ and\ \bibinfo {author} {\bibfnamefont {K.~S.~D.}\ \bibnamefont {Beach}},\ }\bibfield  {title} {\bibinfo {title} {Detection of symmetry-protected topological order in aklt states by exact evaluation of the strange correlator},\ }\href {https://doi.org/10.1103/PhysRevB.93.245141} {\bibfield  {journal} {\bibinfo  {journal} {Phys. Rev. B}\ }\textbf {\bibinfo {volume} {93}},\ \bibinfo {pages} {245141} (\bibinfo {year} {2016}{\natexlab{a}})}\BibitemShut {NoStop}%
\bibitem [{\citenamefont {He}\ \emph {et~al.}(2016)\citenamefont {He}, \citenamefont {Wu}, \citenamefont {You}, \citenamefont {Xu}, \citenamefont {Meng},\ and\ \citenamefont {Lu}}]{Bona2016He}%
  \BibitemOpen
  \bibfield  {author} {\bibinfo {author} {\bibfnamefont {Y.-Y.}\ \bibnamefont {He}}, \bibinfo {author} {\bibfnamefont {H.-Q.}\ \bibnamefont {Wu}}, \bibinfo {author} {\bibfnamefont {Y.-Z.}\ \bibnamefont {You}}, \bibinfo {author} {\bibfnamefont {C.}~\bibnamefont {Xu}}, \bibinfo {author} {\bibfnamefont {Z.~Y.}\ \bibnamefont {Meng}},\ and\ \bibinfo {author} {\bibfnamefont {Z.-Y.}\ \bibnamefont {Lu}},\ }\bibfield  {title} {\bibinfo {title} {Bona fide interaction-driven topological phase transition in correlated symmetry-protected topological states},\ }\href {https://doi.org/10.1103/PhysRevB.93.115150} {\bibfield  {journal} {\bibinfo  {journal} {Phys. Rev. B}\ }\textbf {\bibinfo {volume} {93}},\ \bibinfo {pages} {115150} (\bibinfo {year} {2016})}\BibitemShut {NoStop}%
\bibitem [{\citenamefont {{Zhong}}\ \emph {et~al.}(2017)\citenamefont {{Zhong}}, \citenamefont {{Liu}},\ and\ \citenamefont {{Luo}}}]{Topological2017Zhong}%
  \BibitemOpen
  \bibfield  {author} {\bibinfo {author} {\bibfnamefont {Y.}~\bibnamefont {{Zhong}}}, \bibinfo {author} {\bibfnamefont {Y.}~\bibnamefont {{Liu}}},\ and\ \bibinfo {author} {\bibfnamefont {H.-G.}\ \bibnamefont {{Luo}}},\ }\bibfield  {title} {\bibinfo {title} {{Topological phase in 1D topological Kondo insulator: Z$_{2}$ topological insulator, Haldane-like phase and Kondo breakdown}},\ }\href {https://doi.org/10.1140/epjb/e2017-80102-0} {\bibfield  {journal} {\bibinfo  {journal} {Eur. Phys. J. B}\ }\textbf {\bibinfo {volume} {90}},\ \bibinfo {pages} {147} (\bibinfo {year} {2017})}\BibitemShut {NoStop}%
\bibitem [{\citenamefont {Niu}\ and\ \citenamefont {Qi}(2023)}]{sc23}%
  \BibitemOpen
  \bibfield  {author} {\bibinfo {author} {\bibfnamefont {Y.}~\bibnamefont {Niu}}\ and\ \bibinfo {author} {\bibfnamefont {Y.}~\bibnamefont {Qi}},\ }\bibfield  {title} {\bibinfo {title} {{Strange Correlator for 1D Fermionic Symmetry-Protected Topological Phases}},\ }\href {https://doi.org/10.48550/arXiv.2312.01310} {\bibfield  {journal} {\bibinfo  {journal} {arXiv e-prints}\ ,\ \bibinfo {pages} {arXiv:2312.01310}} (\bibinfo {year} {2023})}\BibitemShut {NoStop}%
\bibitem [{\citenamefont {Lepori}\ \emph {et~al.}(2023)\citenamefont {Lepori}, \citenamefont {Burrello}, \citenamefont {Trombettoni},\ and\ \citenamefont {Paganelli}}]{sc22}%
  \BibitemOpen
  \bibfield  {author} {\bibinfo {author} {\bibfnamefont {L.}~\bibnamefont {Lepori}}, \bibinfo {author} {\bibfnamefont {M.}~\bibnamefont {Burrello}}, \bibinfo {author} {\bibfnamefont {A.}~\bibnamefont {Trombettoni}},\ and\ \bibinfo {author} {\bibfnamefont {S.}~\bibnamefont {Paganelli}},\ }\bibfield  {title} {\bibinfo {title} {{Strange correlators for topological quantum systems from bulk-boundary correspondence}},\ }\href {https://doi.org/10.1103/PhysRevB.108.035110} {\bibfield  {journal} {\bibinfo  {journal} {Phys. Rev. B}\ }\textbf {\bibinfo {volume} {108}},\ \bibinfo {pages} {035110} (\bibinfo {year} {2023})}\BibitemShut {NoStop}%
\bibitem [{\citenamefont {{Zhang}}\ \emph {et~al.}(2022)\citenamefont {{Zhang}}, \citenamefont {{Qi}},\ and\ \citenamefont {{Bi}}}]{sc21}%
  \BibitemOpen
  \bibfield  {author} {\bibinfo {author} {\bibfnamefont {J.-H.}\ \bibnamefont {{Zhang}}}, \bibinfo {author} {\bibfnamefont {Y.}~\bibnamefont {{Qi}}},\ and\ \bibinfo {author} {\bibfnamefont {Z.}~\bibnamefont {{Bi}}},\ }\bibfield  {title} {\bibinfo {title} {{Fidelity Strange Correlators for Average Symmetry-Protected Topological Phases}},\ }\href {https://doi.org/10.48550/arXiv.2210.17485} {\bibfield  {journal} {\bibinfo  {journal} {arXiv e-prints}\ ,\ \bibinfo {eid} {arXiv:2210.17485}} (\bibinfo {year} {2022})},\ \Eprint {https://arxiv.org/abs/2210.17485} {arXiv:2210.17485 [cond-mat.str-el]} \BibitemShut {NoStop}%
\bibitem [{\citenamefont {Zhou}\ \emph {et~al.}(2022{\natexlab{a}})\citenamefont {Zhou}, \citenamefont {Li}, \citenamefont {Yan}, \citenamefont {Ye},\ and\ \citenamefont {Meng}}]{sc26}%
  \BibitemOpen
  \bibfield  {author} {\bibinfo {author} {\bibfnamefont {C.}~\bibnamefont {Zhou}}, \bibinfo {author} {\bibfnamefont {M.-Y.}\ \bibnamefont {Li}}, \bibinfo {author} {\bibfnamefont {Z.}~\bibnamefont {Yan}}, \bibinfo {author} {\bibfnamefont {P.}~\bibnamefont {Ye}},\ and\ \bibinfo {author} {\bibfnamefont {Z.~Y.}\ \bibnamefont {Meng}},\ }\bibfield  {title} {\bibinfo {title} {Detecting subsystem symmetry protected topological order through strange correlators},\ }\href {https://doi.org/10.1103/PhysRevB.106.214428} {\bibfield  {journal} {\bibinfo  {journal} {Phys. Rev. B}\ }\textbf {\bibinfo {volume} {106}},\ \bibinfo {pages} {214428} (\bibinfo {year} {2022}{\natexlab{a}})}\BibitemShut {NoStop}%
\bibitem [{\citenamefont {Zhang}\ \emph {et~al.}(2024)\citenamefont {Zhang}, \citenamefont {Li},\ and\ \citenamefont {Ye}}]{sc25}%
  \BibitemOpen
  \bibfield  {author} {\bibinfo {author} {\bibfnamefont {J.-Y.}\ \bibnamefont {Zhang}}, \bibinfo {author} {\bibfnamefont {M.-Y.}\ \bibnamefont {Li}},\ and\ \bibinfo {author} {\bibfnamefont {P.}~\bibnamefont {Ye}},\ }\bibfield  {title} {\bibinfo {title} {{Higher-Order Cellular Automata Generated Symmetry-Protected Topological Phases and Detection Through Multi Point Strange Correlators}},\ }\href {https://doi.org/10.1103/PRXQuantum.5.030342} {\bibfield  {journal} {\bibinfo  {journal} {PRX Quantum}\ }\textbf {\bibinfo {volume} {5}},\ \bibinfo {pages} {030342} (\bibinfo {year} {2024})}\BibitemShut {NoStop}%
\bibitem [{\citenamefont {Zhou}\ \emph {et~al.}(2022{\natexlab{b}})\citenamefont {Zhou}, \citenamefont {Li}, \citenamefont {Yan}, \citenamefont {Ye},\ and\ \citenamefont {Meng}}]{sc24}%
  \BibitemOpen
  \bibfield  {author} {\bibinfo {author} {\bibfnamefont {C.}~\bibnamefont {Zhou}}, \bibinfo {author} {\bibfnamefont {M.-Y.}\ \bibnamefont {Li}}, \bibinfo {author} {\bibfnamefont {Z.}~\bibnamefont {Yan}}, \bibinfo {author} {\bibfnamefont {P.}~\bibnamefont {Ye}},\ and\ \bibinfo {author} {\bibfnamefont {Z.~Y.}\ \bibnamefont {Meng}},\ }\bibfield  {title} {\bibinfo {title} {{Detecting subsystem symmetry protected topological order through strange correlators}},\ }\href {https://doi.org/10.1103/PhysRevB.106.214428} {\bibfield  {journal} {\bibinfo  {journal} {Phys. Rev. B}\ }\textbf {\bibinfo {volume} {106}},\ \bibinfo {pages} {214428} (\bibinfo {year} {2022}{\natexlab{b}})}\BibitemShut {NoStop}%
\bibitem [{\citenamefont {Penrose}(1955)}]{Penrose_1955}%
  \BibitemOpen
  \bibfield  {author} {\bibinfo {author} {\bibfnamefont {R.}~\bibnamefont {Penrose}},\ }\bibfield  {title} {\bibinfo {title} {A generalized inverse for matrices},\ }\href {https://doi.org/10.1017/S0305004100030401} {\bibfield  {journal} {\bibinfo  {journal} {Mathematical Proceedings of the Cambridge Philosophical Society}\ }\textbf {\bibinfo {volume} {51}},\ \bibinfo {pages} {406–413} (\bibinfo {year} {1955})}\BibitemShut {NoStop}%
\bibitem [{\citenamefont {Fannes}\ \emph {et~al.}(1992{\natexlab{b}})\citenamefont {Fannes}, \citenamefont {Nachtergaele},\ and\ \citenamefont {Werner}}]{Fannes1992}%
  \BibitemOpen
  \bibfield  {author} {\bibinfo {author} {\bibfnamefont {M.}~\bibnamefont {Fannes}}, \bibinfo {author} {\bibfnamefont {B.}~\bibnamefont {Nachtergaele}},\ and\ \bibinfo {author} {\bibfnamefont {R.~F.}\ \bibnamefont {Werner}},\ }\bibfield  {title} {\bibinfo {title} {Finitely correlated states on quantum spin chains},\ }\href {https://doi.org/10.1007/BF02099178} {\bibfield  {journal} {\bibinfo  {journal} {Communications in Mathematical Physics}\ }\textbf {\bibinfo {volume} {144}},\ \bibinfo {pages} {443} (\bibinfo {year} {1992}{\natexlab{b}})}\BibitemShut {NoStop}%
\bibitem [{\citenamefont {O'Brien}\ \emph {et~al.}(2020)\citenamefont {O'Brien}, \citenamefont {Vernier},\ and\ \citenamefont {Fendley}}]{OBrien2020}%
  \BibitemOpen
  \bibfield  {author} {\bibinfo {author} {\bibfnamefont {E.}~\bibnamefont {O'Brien}}, \bibinfo {author} {\bibfnamefont {E.}~\bibnamefont {Vernier}},\ and\ \bibinfo {author} {\bibfnamefont {P.}~\bibnamefont {Fendley}},\ }\bibfield  {title} {\bibinfo {title} {``not-$a$'', representation symmetry-protected topological, and potts phases in an ${S}_{3}$-invariant chain},\ }\href {https://doi.org/10.1103/PhysRevB.101.235108} {\bibfield  {journal} {\bibinfo  {journal} {Phys. Rev. B}\ }\textbf {\bibinfo {volume} {101}},\ \bibinfo {pages} {235108} (\bibinfo {year} {2020})}\BibitemShut {NoStop}%
\bibitem [{\citenamefont {Wierschem}\ and\ \citenamefont {Beach}(2016{\natexlab{b}})}]{Hspin2}%
  \BibitemOpen
  \bibfield  {author} {\bibinfo {author} {\bibfnamefont {K.}~\bibnamefont {Wierschem}}\ and\ \bibinfo {author} {\bibfnamefont {K.~S.~D.}\ \bibnamefont {Beach}},\ }\bibfield  {title} {\bibinfo {title} {Detection of symmetry-protected topological order in aklt states by exact evaluation of the strange correlator},\ }\href {https://doi.org/10.1103/PhysRevB.93.245141} {\bibfield  {journal} {\bibinfo  {journal} {Phys. Rev. B}\ }\textbf {\bibinfo {volume} {93}},\ \bibinfo {pages} {245141} (\bibinfo {year} {2016}{\natexlab{b}})}\BibitemShut {NoStop}%
\bibitem [{\citenamefont {Giuli}(1942)}]{Racah}%
  \BibitemOpen
  \bibfield  {author} {\bibinfo {author} {\bibfnamefont {R.}~\bibnamefont {Giuli}},\ }\bibfield  {title} {\bibinfo {title} {{Theory of Complex Spectra. II}},\ }\href {https://doi.org/10.1103/PhysRev.62.438} {\bibfield  {journal} {\bibinfo  {journal} {Phys. Rev. Lett.}\ }\textbf {\bibinfo {volume} {62}},\ \bibinfo {pages} {438} (\bibinfo {year} {1942})}\BibitemShut {NoStop}%
\bibitem [{\citenamefont {Rudin}(1976)}]{rudin1976principles}%
  \BibitemOpen
  \bibfield  {author} {\bibinfo {author} {\bibfnamefont {W.}~\bibnamefont {Rudin}},\ }\href {https://books.google.com/books?id=kwqzPAAACAAJ} {\emph {\bibinfo {title} {Principles of Mathematical Analysis}}},\ International series in pure and applied mathematics\ (\bibinfo  {publisher} {McGraw-Hill},\ \bibinfo {year} {1976})\BibitemShut {NoStop}%
\bibitem [{\citenamefont {Mityagin}(2015)}]{mityagin2015zerosetrealanalytic}%
  \BibitemOpen
  \bibfield  {author} {\bibinfo {author} {\bibfnamefont {B.}~\bibnamefont {Mityagin}},\ }\href {https://arxiv.org/abs/1512.07276} {\bibinfo {title} {The zero set of a real analytic function}} (\bibinfo {year} {2015}),\ \Eprint {https://arxiv.org/abs/1512.07276} {arXiv:1512.07276 [math.CA]} \BibitemShut {NoStop}%
\end{thebibliography}
\end{document}